\documentclass[aps,prd,nobibnotes,twocolumn,superscriptaddress,bibliography]{revtex4-1}

\usepackage{amsfonts}
\usepackage{mathrsfs}
\usepackage{amsmath}
\usepackage{color}
\usepackage{graphicx}
\usepackage{bm}
\usepackage{amssymb}
\usepackage{xspace}
\usepackage{epstopdf}
\usepackage{dcolumn}
\usepackage{longtable}
\usepackage{multirow}
\usepackage{float}
\usepackage{comment}
\usepackage{soul}

\usepackage[colorlinks=true, letterpaper=true, pdfstartview=FitV, linkcolor=blue, citecolor=blue, urlcolor=blue]{hyperref}

\makeatother
\begin{document}
	\title{Magnetic higher-order nodal lines }

	\author{Zeying Zhang}
	\affiliation{College of Mathematics and Physics, Beijing University of Chemical Technology, Beijing 100029, China}

	\author{Zhi-Ming Yu}
    \email{zhiming\_yu@bit.edu.cn}
	\address{Key Lab of Advanced Optoelectronic Quantum Architecture and Measurement (MOE), Beijing Key Lab of Nanophotonics \& Ultrafine Optoelectronic Systems, and School of Physics, Beijing Institute of Technology, Beijing 100081, China}

    \author{Shengyuan A. Yang}
\address{Research Laboratory for Quantum Materials, Singapore University of
	Technology and Design, Singapore 487372, Singapore}	

\begin{abstract}
Nodal lines, as one-dimensional band degeneracies in momentum space, usually feature a linear energy splitting. Here, we propose the concept of magnetic higher-order nodal lines, which are nodal lines with higher-order energy splitting and realized in magnetic systems with broken time reversal symmetry. We provide sufficient symmetry conditions for stabilizing magnetic quadratic and cubic nodal lines, based on which concrete lattice models are constructed to demonstrate their existence. Unlike its counterpart in nonmagnetic systems, the magnetic quadratic nodal line can exist as the only band degeneracy at the Fermi level. We show that these nodal lines can be accompanied by torus surface states, which form a surface band that span over the whole surface Brillouin zone. Under symmetry breaking, these magnetic nodal lines can be transformed into a variety of interesting topological states, such as three-dimensional quantum anomalous Hall insulator, multiple linear nodal lines, and magnetic triple-Weyl semimetal. The three-dimensional quantum anomalous Hall insulator features a Hall conductivity $\sigma_{xy}$ quantized in unit of $e^2/(h d)$ where $d$ is the lattice constant normal to the $x$-$y$ plane. Our work reveals previously unknown topological states, and offers guidance to search for them in realistic material systems.
\end{abstract}
\maketitle
	
\section{Introduction }

Topological metals and semimetals have been attracting significant interest in current
research \cite{chiu_classification_2016,armitage_weyl_2018,yang_dirac_2016,Burkov2016Topological-Nm}. These states are characterized by the symmetry/topology protected band degeneracies near the Fermi level, because they determine the low-energy quasiparticle excitations and hence the physical properties of the system~\cite{Nielsen1983Adler-PLB,Volovik2003universe-OUPoD,HosurPRB-2012,Son2013Chiral-PRB,Quantized_NC_2017,LiuY-PRL2020}. Therefore, a central task in the field is to discover and classify all possible types of protected band degeneracies.

Such degeneracies can be classified from different perspectives. For example, regarding the dimensionality of the degeneracy manifold in momentum space, the band degeneracies
can be classified into zero-dimensional (0D) nodal points~\cite{wan_topological_2011,Young2012Dirac-Prl, wang_dirac_2012,Zhao2013, bradlyn_beyond_2016}, 1D nodal lines~\cite{burkov_topological_2011, yang_dirac_2014,weng_topological_2015,Chen2015,Yu2015,Kim2015Dirac-Prl,MullenPRL_2015}, or even 2D nodal surfaces  \cite{Liang2016Node-PRB,Zhong_NS_2016,wu_nodal_2018,turker_weyl_2018,Zhang2018Nodal-PRB}. For each class, further classification can be made based on the number of degeneracy, the resulting Fermi surface topology, the distribution in the Brillouin zone (BZ), and etc.

In the classification scheme, the character of band dispersion clearly plays an important role, as it directly affects the density of states, the group velocity, and possible topological charge of the low-energy quasiparticles. For most cases, the dispersion around a band degeneracy is of linear type, namely, the degeneracy is formed by the linear crossing between two bands. Nevertheless, under certain symmetries, the linear order term may be forbidden, and then the leading order dispersion in the band energy splitting will be pushed to higher orders \cite{Xu2011Chern-Prl,fang_multi-weyl_2012,liu_predicted_2017, he_observation_2020,yang_classification_2014}. For example, it was found that there exist symmetry-protected twofold nodal points, known as multi-Weyl points, around which the leading order band splitting is quadratic or even cubic along certain directions \cite{fang_multi-weyl_2012}. Similar study was later extended to fourfold Dirac points \cite{yang_classification_2014,Gao2016Classification-PRB,Yu2018Nonsymmorphic-PRM}, and a systematic classification of higher-order Dirac points was achieved in Ref.~\cite{wu_higher-order_2020}.

Recently, in Ref.~\cite{yu_quadratic_2019}, Yu \emph{et al.} discovered the possibility of higher-order nodal lines in nonmagnetic systems. They found that quadratic or cubic dispersion could be the leading order dispersion of band splitting in the transverse plane for every point on the line.  The special dispersion leads to many interesting effects, including distinct scalings in the thermodynamic and response properties, unusual surface states, and rich topological phases resulted from symmetry breaking \cite{yu_quadratic_2019,wang_possible_2020}.

The study of Ref.~\cite{yu_quadratic_2019} is restricted to nonmagnetic systems which preserve the time reversal symmetry $\mathcal{T}$. For magnetic systems, $\mathcal{T}$ is broken, which will fundamentally impact the topological classification. A trend in recent research is to push the study of topological phases to magnetic systems \cite{morali2019fermi,You2019,liu2019magnetic,belopolski2019discovery,xu2020high,Jin_double_PRB2020,Mag_Puphal_PRL2020,Mag_XGang_npj2020}. Particularly, the magnetic linear nodal lines have been actively explored in magnetic materials \cite{wang_antiferromagnetic_2017,Wang2018,Wang2019,Chen2019,Feng2019,nie_magnetic_2020,Hezhang_NL_PRB2020,song_symmetry-protected_2020,ZhangRW_NL_2020}.

In view of these recent advances, a natural question is: \emph{Is it possible to have protected magnetic higher-order nodal lines?}

In this work, we answer the above question in the affirmative. We show that nodal lines with quadratic (cubic) leading order dispersion can be realized in magnetic systems with spin-orbit coupling fully considered, which we term as magnetic quadratic (cubic) nodal lines. We provide sufficient symmetry conditions for their protection, based on which we construct lattice models to explicitly demonstrate the existence of these nodal lines. Notably, unlike nonmagnetic systems, we find that magnetic systems can host quadratic nodal lines as the only band degeneracy at Fermi level. Both quadratic and cubic nodal lines may be accompanied by a special kind of torus surface states, which span over the whole surface BZ and lead to a large surface density of states. Furthermore, under symmetry breaking, these magnetic nodal lines can evolve into a variety of intriguing topological states, such as the 3D quantum anomalous Hall (QAH) insulator, multiple linear nodal lines, and the magnetic triple-Weyl semimetal. Particularly, the 3D QAH insulator features a Hall conductivity $\sigma_{xy}$ quantized in unit of $e^2/(hd)$, with $d$ the lattice constant normal to the $x$-$y$ plane. Our work reveals previously unknown topological phases in systems with $\mathcal{T}$ breaking, offers detailed guidance to search for them in real material systems, and will stimulate further studies on their fascinating properties.

\section{Magnetic quadratic nodal line} \label{MQNL}

\subsection{Symmetry condition}

\begin{figure}[t]
	\includegraphics[width=\columnwidth]{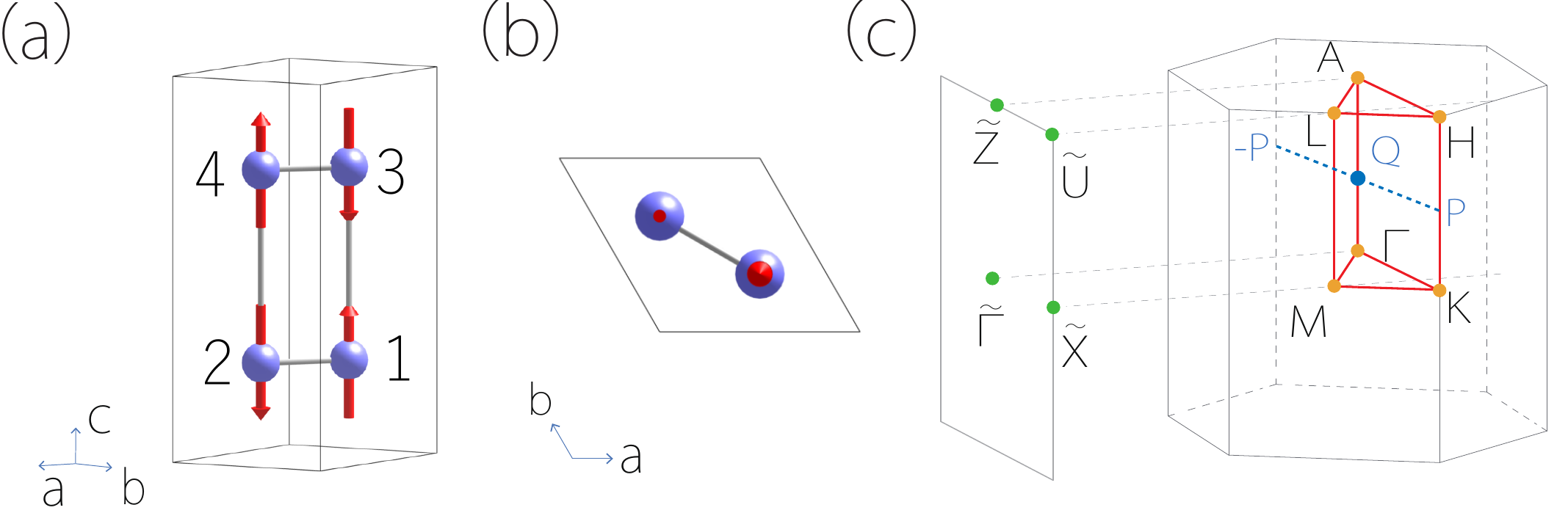}
	\caption{
		(a) Side view and (b) top view of the unit cell for the MQNL model. The red arrows denote the magnetic moments on the sites. 
		(c) Bulk and surface BZ of the lattice model.	}
	\label{fig:QNL_BZ}
\end{figure}

Let's first present a symmetry condition that can protect a magnetic quadratic nodal line (MQNL) in magnetic systems. The symmetries should help to protect the degeneracy along a 1D line and also eliminate the linear term in the band splitting around this line. Here, we find that a MQNL can be stabilized on a high-symmetry line that is an invariant subspace of two symmetries: a three fold rotation $C_{3z}$ and a magnetic symmetry $M_{z}{\cal T}$ involving a mirror, where we take the rotation axis to be along the $z$ direction. Here, $M_{z}{\cal T}$ is a combined symmetry, while the individual $M_z$ and $\mathcal{T}$ symmetries are broken by the magnetism. The two symmetries restrict the MQNL to the $\Gamma$-$A$ path of the hexagonal BZ, as shown in Fig.~\ref{fig:QNL_BZ}(c).

To explicitly demonstrate the quadratic dispersion, we take an arbitrary point $Q$ on the $\Gamma$-$A$ path. The Bloch states at $Q$ can be chosen as the eigenstates of $C_{3z}$. For magnetic systems, it is necessary to consider the spin-orbit coupling. Then the  $C_{3z}$ eigenvalues are given by
$c_{3z}=-1,\ e^{\pm i\pi/3}$, and we denote the eigenstates as $|c_{3z}\rangle$ by using the eigenvalues.  Since $C_{3z}$ commutes with $M_{z}{\cal T}$, we have
\begin{eqnarray}
C_{3z}\left(M_{z}{\cal T}|c_{3z}\rangle\right) & = & c_{3z}^{*}\left(M_{z}{\cal T}|c_{3z}\rangle\right),
\end{eqnarray}
which indicates that the two states $|e^{i\pi/3}\rangle$ and $|e^{-i\pi/3}\rangle$ must always form a pair, degenerate in energy.  Hence, this degeneracy leads to a doubly degenerate nodal line along the $\Gamma$-$A$ path. In the basis of these two degenerate states, the matrix representations of the symmetry operators are given by
\begin{eqnarray}
C_{3z}=e^{i\sigma_{3}\pi/3},\ \  &  & { M}_{z}{\cal T}=\sigma_{1}{\cal K},
\end{eqnarray}
with ${\sigma}_i$'s the Pauli matrices and ${\cal K}$ the complex conjugation operator.
Then, the effective Hamiltonian constrained by  $C_{3z}$ and ${ M}_{z}{\cal T}$ around $Q$ in the transverse plane can be obtained as
\begin{eqnarray}
{\cal H}_{\text{eff}}^Q(\boldsymbol{k}) & = & c k_{\|}^2+\alpha k_{-}^{2}\sigma_{+}+\alpha^{*}k_{+}^{2}\sigma_{-},\label{eq:kpham}
\end{eqnarray}
where $k_{\|}=\sqrt{k_x^2+k_y^2}$, $c$  ($\alpha$) is a real (complex) model parameter which generally depends on $Q$, $k_{\pm}=k_{x}\pm ik_{y}$, and $\sigma_{\pm}=(\sigma_{1}\pm i\sigma_{2})/2$.
The spectrum of this Hamiltonian (\ref{eq:kpham}) is
\begin{eqnarray}
E & = &(c\pm |\alpha|) k_{\|}^2,\label{eq:kpegqnl}
\end{eqnarray}
which confirms that the leading order dispersion in the band splitting is quadratic and thus the nodal line here is indeed a MQNL.

Before proceeding, we note that although the presence of the above MQNL does not require a vertical mirror symmetry (denoted as $M_x$ without loss of generality), MQNLs are compatible with this symmetry such as in the lattice model to be discussed in a while. When we do have this symmetry, its matrix representation in the basis of the degenerate pair will be given by $M_{x}=-i\sigma_{1}$, and one can show that the form of the effective Hamiltonian (\ref{eq:kpham}) remains unchanged, but $\alpha$ will be constrained to be a real number.

\subsection{Lattice model for MQNL}

We construct a lattice model with the symmetry condition specified above to explicitly demonstrate the existence of the MQNL.
Consider a 3D lattice consisting of 2D honeycomb lattices AA-stacked along the $z$ direction, as shown in Fig.~\ref{fig:QNL_BZ}(a-b).
Here, each unit cell contains two layers and four active sites. The four sites are labeled as 1 to 4, as in Fig.~\ref{fig:QNL_BZ}(a). At each site, we put one $s$-like basis orbital $\phi_s$ with a specific spin polarization, such that the basis of our model in a unit cell is given by $\Phi=(\phi_s(1)|\uparrow\rangle, \phi_s(2)|\downarrow\rangle, \phi_s(3)|\downarrow\rangle, \phi_s(4)|\uparrow\rangle)$. Physically, this case may correspond to a G-type antiferromagnetic (AFM) order as illustrated in Fig.~\ref{fig:QNL_BZ}(a).

We require that the model respects the $C_{3z}$ and $M_{z}{\cal T}$ symmetries. In the basis of $\Phi$, these symmetry operators take the form of
\begin{eqnarray}
C_{3z}=e^{i\Gamma_{33}\pi/3},\qquad & M_{z}{\cal T}=i\Gamma_{10}{\cal K},
\label{eq:qnl1}
\end{eqnarray}
where we define $\Gamma_{\mu\nu}\equiv\sigma_{\mu}\otimes\sigma_{\nu}$ and $\sigma_{0}$ denotes the $2\times2$ identity matrix. In addition, for convenience of later discussion, we also impose a vertical mirror $M_x$, which interchanges A and B sites in each layer, such that it is represented by
\begin{equation}
  M_x=i\Gamma_{01}.
\label{eq:qnl2}
\end{equation}

Following the standard approach as in Refs.~\cite{wieder_spin-orbit_2016,Yu2019Circumventing-PRB}, we can construct the lattice model that satisfies the symmetries in Eqs.~(\ref{eq:qnl1}) and (\ref{eq:qnl2}). In momentum space, the obtained model can be written as
\begin{widetext}
  \begin{eqnarray}
&  & H=A_{0}-A_{1}\sin\frac{k_{y}}{2}\left(\sin\frac{k_{x}}{2\sqrt{3}}\Gamma_{32}+\cos\frac{k_{x}}{2\sqrt{3}}\Gamma_{31}\right) +\frac{A_{1}}{\sqrt{3}}\left[\left(\sin\frac{k_{x}}{\sqrt{3}}+\sin\frac{k_{x}}{2\sqrt{3}}\cos\frac{k_{y}}{2}\right)\Gamma_{02}\right.\nonumber\\
&  &\qquad-\left(\left.\cos\frac{k_{x}}{\sqrt{3}}-\cos\frac{k_{x}}{2\sqrt{3}}\cos\frac{k_{y}}{2}\right)\Gamma_{01}\right]+\left(\cos\frac{k_{x}}{\sqrt{3}}
+2\cos\frac{k_{x}}{2\sqrt{3}}\cos\frac{k_{y}}{2}\right)\left(A_{2}\cos\frac{k_{z}}{2}\Gamma_{11}+A_{3}\sin\frac{k_{z}}{2}\Gamma_{21}\right)\nonumber \\
&  & \qquad -\left(\sin\frac{k_{x}}{\sqrt{3}}-2\sin\frac{k_{x}}{2\sqrt{3}}\cos\frac{k_{y}}{2}\right)\left(A_{2}\cos\frac{k_{z}}{2}\Gamma_{12}+A_{3}\sin\frac{k_{z}}{2}\Gamma_{22}\right).
\label{eq:TBham}
\end{eqnarray}
\end{widetext}
Here, $A_{i}$ ($i=0,1,2,3$) are real model parameters. One can readily check that the Hamiltonian (\ref{eq:TBham}) is invariant under the symmetry operators in Eqs.~(\ref{eq:qnl1}) and (\ref{eq:qnl2}).

\begin{figure}[b]
	\includegraphics[width=\columnwidth]{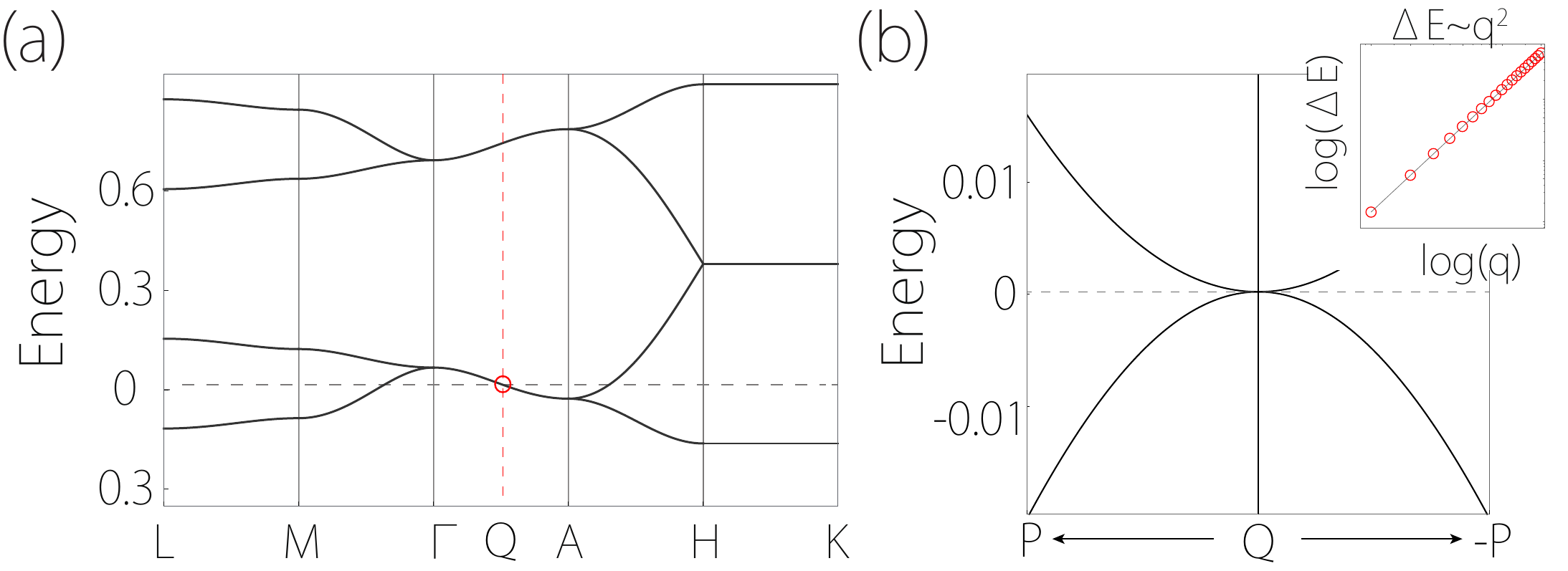}
	\caption{(a) Band structure of the lattice model (\ref{eq:TBham}) along high-symmetry lines.	(b)  Enlarged view of the band structure along $P$-$Q$ path [marked in Fig.~\ref{fig:QNL_BZ}(c)] which lies in the plane perpendicular to $k_z$. The inset shows the log-log plot for the band splitting $\Delta E$ versus the momentum deviation from the nodal line. In the model, we set $A_0= 0.35$, $A_1 = 0.3$, $A_2 = 0.1$, and $A_3 = 0.13$.	}
	\label{fig:QNL_band}
\end{figure}

The calculated band structure of this model is plotted in Fig.~\ref{fig:QNL_band}(a). One observes that both lower two bands and upper two bands form twofold nodal lines along the $\Gamma$-$A$ path. One can check that these nodal lines are indeed MQNLs. In Fig.~\ref{fig:QNL_band}(b), we plot the dispersion around a generic point $Q$ [marked in Fig.~\ref{fig:QNL_band}(a)] on the lower nodal line, which confirms the quadratic band splitting.

In fact, one can expand the model (\ref{eq:TBham}) around point $Q$ to obtain a two-band effective model in the plane transverse to the nodal line. As shown in Appendix~\ref{appendix1}, the result recovers the effective model in Eq.~(\ref{eq:kpham}), confirming that this is indeed the desired MQNL.

It should be pointed out that if we focus on the lower two bands in Fig.~\ref{fig:QNL_band}(a), the MQNL on $\Gamma$-$A$ is their only degeneracy in the BZ. When the electron filling has only the lowest band filled, i.e., the Fermi level is around the nodal line, then the quasiparticles around the MQNL will play a dominant role in the physical properties of the system. In comparison, in nonmagnetic systems, the two bands that form a quadratic nodal line must also degenerate on other paths (such as the nodal line lying in $\Gamma MK$ plane) in the BZ, due to the presence of $M_z$~\cite{yu_quadratic_2019}. In other words, the quadratic nodal line in nonmagnetic systems must be accompanied by other degeneracies in the BZ, whereas it can be the only one for magnetic systems.

\section{Magnetic cubic nodal line}
\subsection{Symmetry condition}

Next, we investigate the possibility of cubic nodal lines in magnetic systems. In the following, we show that magnetic cubic nodal lines (MCNLs) can be protected in the invariant subspace of two symmetries: a sixfold rotation $C_{6z}$ and a vertical mirror $M_x$ that contains the rotation axis. Hence, in the BZ, such MCNL can only appear on the $\Gamma$-$A$ path. Note that these symmetries do not involve the time reversal, so the symmetry condition here actually applies for both nonmagnetic and magnetic systems.

\begin{figure}[b]
	\includegraphics[width=\columnwidth]{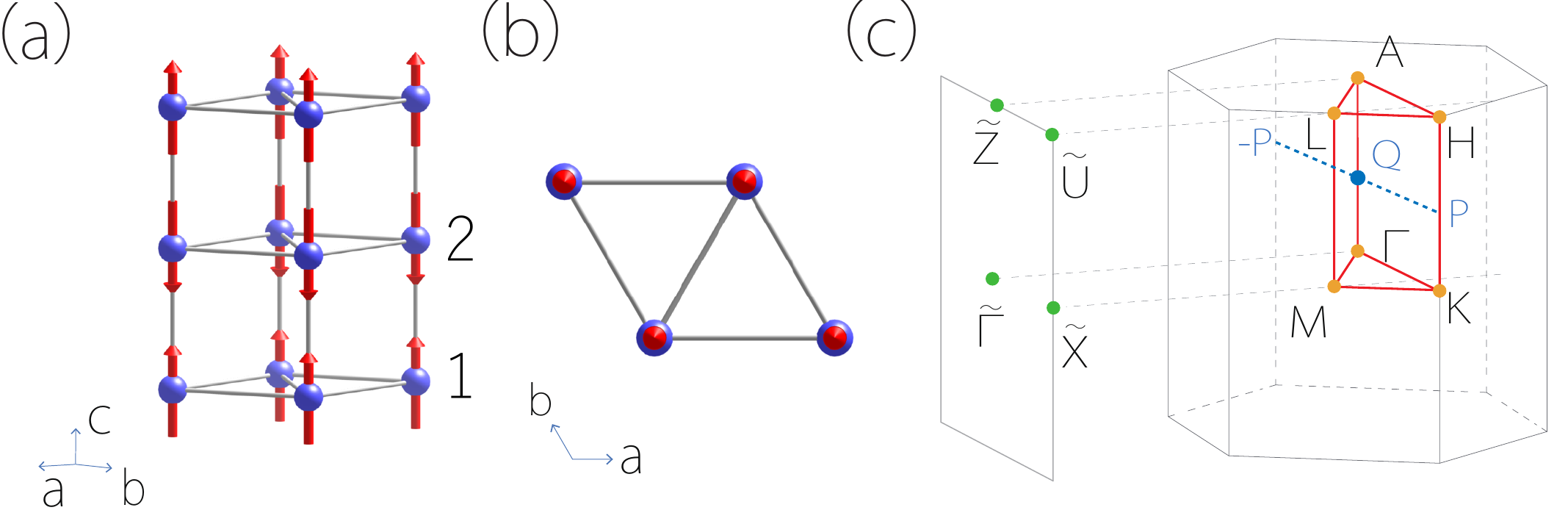}
	\caption{(a) Side view and (b) top view of the unit cell for the MCNL model. The directions of local moments are denoted by the red arrows. (c) Bulk and surface BZ of the lattice model.}
	\label{fig:CNL_BZ}
\end{figure}

Let's consider a generic point $Q$ on the high-symmetry path $\Gamma$-$A$, which has both $C_{6z}$ and $M_x$ symmetries. The Bloch states at $Q$ can be chosen as the eigenstates of $C_{6z}$. We may denote these eigenstates $|c_{6z}\rangle$ by the  $C_{6z}$ eigenvalues, with $c_{6z}=\pm i,\ e^{\pm i\pi/6},\ -e^{\pm i\pi/6}$. We have the following relations between $C_{6z}$ and $M_x$,
\begin{eqnarray}
C_{6z}M_x & = & M_x C_{6z}^{-1},
\end{eqnarray}
so that
\begin{eqnarray}
C_{6z}\left(M_x|c_{6z}\rangle\right) & = & M_x\left(C_{6z}^{-1}|c_{6z}\rangle\right)=c_{6z}^*\left(M_x|c_{6z}\rangle\right).
\end{eqnarray}
Hence, each state $|c_{6z}\rangle$ at $Q$ must have a degenerate partner with eigenvalue $c_{6z}^*$, which leads to twofold nodal lines on the $\Gamma$-$A$ path. And all these $|c_{6z}\rangle$'s form three degenerate pairs: $(|i\rangle, |-i\rangle)$, $(|\varepsilon\rangle, |\varepsilon^*\rangle)$, and $(|-\varepsilon\rangle, |-\varepsilon^*\rangle)$, where $\varepsilon=e^{i\pi/6}$.

The MCNL corresponds to the degenerate pair with eigenvalues $c_{6z}=\pm i$. In the two-state basis $(|i\rangle, |-i\rangle)$, the matrix representation of the symmetry operators can be written as
\begin{eqnarray}
C_{6z}=i\sigma_{3},\qquad  &  & M_x=i\sigma_{1}.
\label{eq:cnl1}
\end{eqnarray}
Then the effective Hamiltonian at $Q$ constrained by these symmetries in the plane transverse to the nodal line is obtained as
\begin{eqnarray}
{\cal H}_{\text{eff}}^{Q}(\boldsymbol{k}) & = & c_1 k_{\|}^2+[i(c_2 k_{-}^{3}+c_3 k_{+}^{3})\sigma_{+}+h.c.],\label{eq:CNLkpham}
\end{eqnarray}
where  $c_i$'s   ($i=1, 2, 3$) are real model parameter. Note that the first (quadratic) term is proportional to the identity matrix, so it does not affect the leading order of the band splitting.
Indeed, the spectrum of (\ref{eq:CNLkpham}) is
\begin{eqnarray}
E & = &c_1 k_{\|}^2\pm |(c_2 k_{-}^{3}+c_3 k_{+}^{3})|,\label{eq:kpegcnl}
\end{eqnarray}
showing that leading order in the band splitting is cubic. Thus, this nodal line on the $\Gamma$-$A$ path is a MCNL.

\subsection{Lattice model for MCNL}
\label{LatticeModelForMCNL}
Guided by the above symmetry condition, we construct a concrete lattice model to demonstrate the existence of the MCNL. Consider the 3D lattice as illustrated in Fig.~\ref{fig:CNL_BZ}(a-b), formed by stacking 2D triangular lattices along the $z$ direction. Let's consider the A-type AFM ordering as in Fig.~\ref{fig:CNL_BZ}(a). This preserves the $C_{6z}$ symmetry and also a glide mirror symmetry  $\tilde{M}_x=\{M_x|00\frac{1}{2}\}$. Note that the glide character of $\tilde{M}_x$ is not essential for the MCNL. Indeed, the symmetry analysis in the previous section is not affected by the fractional translation of the glide mirror.

\begin{figure}[t]
	\includegraphics[width=\columnwidth]{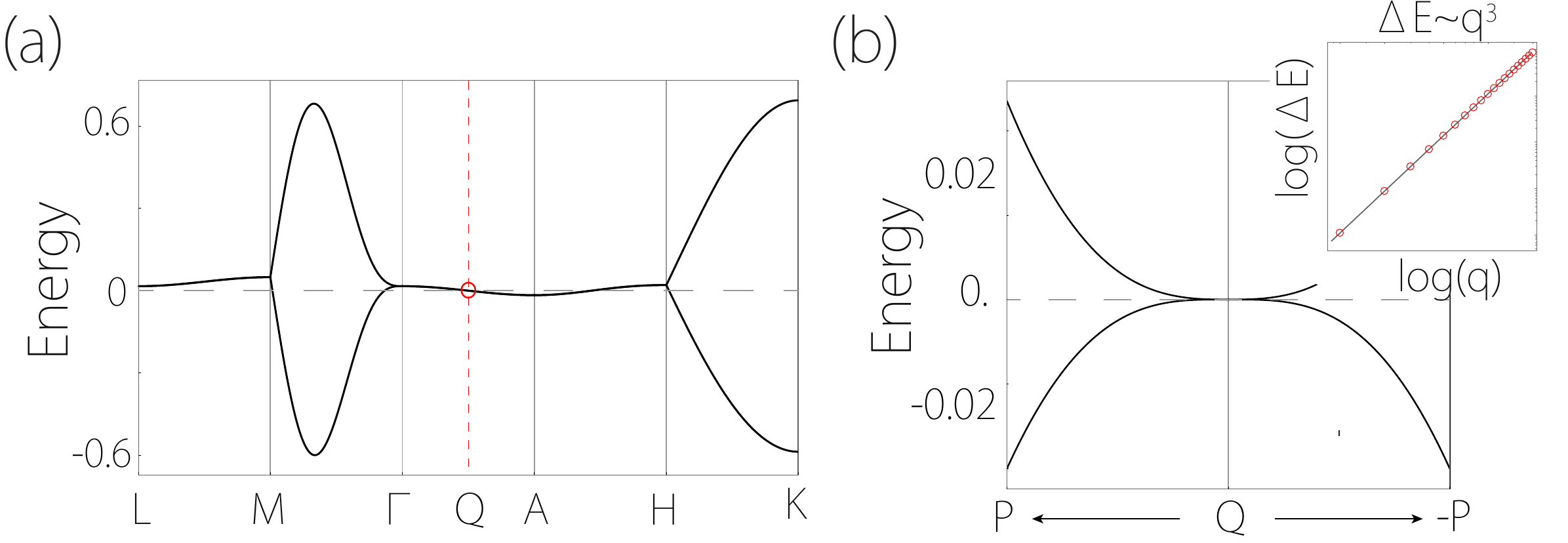}
	\caption{(a) Band structure of the lattice model (\ref{eq:CTBham}) along high-symmetry lines.	(b)  Enlarged view of the band structure along the $P$-$Q$ path [marked in Fig.~\ref{fig:CNL_BZ}(c)] which is in the plane perpendicular to $k_z$. The inset shows the log-log plot for the band splitting $\Delta E$ versus the momentum deviation from the nodal line. In the model,  we take   $A_0=0.03$, $A_1=0.02$, $A_2= -0.01$ $A_3 = 0.3$, and $A_4 = 0.3$.}
	\label{fig:CNL_band}
\end{figure}

In this lattice model, a unit cell contains two sites, labeled as 1 and 2 as in Fig.~\ref{fig:CNL_BZ}(a). We put one basis orbital on each site: at site 1, it is the $p_+$ orbital with spin up; and at site 2, it is the $p_-$ orbital with spin down. In other words, the basis of our model is taken to be $\Phi=(\phi_{p_+}(1)|\uparrow\rangle,\phi_{p_-}(2)|\downarrow\rangle)$.
This setup conforms with the specified symmetries above. In this basis, the symmetry operators take the form of
\begin{eqnarray}\label{CNL-Latt}
C_{6z}=i \sigma_{3},\qquad   \tilde{M}_x=i \sigma_{1},
\end{eqnarray}
which are the same as in Eq.~(\ref{eq:cnl1}).
Constrained by these symmetries, the obtained lattice model in momentum space can be expressed as
\begin{widetext}
\begin{eqnarray}
&  & H=A_0+A_1 \cos k_z+ A_2 (2\cos \frac{\sqrt{3} k_x}{2} \cos \frac{k_y}{2}+ \cos k_y)+ A_3\cos \frac{k_z}{2}\left( \sin k_y -2  \cos \frac{\sqrt{3} k_x}{2} \sin \frac{k_y}{2}\right)\sigma _1 \nonumber \\
&  &\qquad+ A_4\cos \frac{k_z}{2}\left[2  \sin \frac{\sqrt{3} k_x}{2} \cos \frac{3 k_y}{2} - \sin (\sqrt{3} k_x) \right]\sigma _2.
\label{eq:CTBham}
\end{eqnarray}
\end{widetext}
Again, the $A_i$'s here are real model parameters. One can readily check that the Hamiltonian (\ref{eq:CTBham}) is invariant under the symmetry operators in Eq.~(\ref{CNL-Latt}).

The calculated band structure of this model (\ref{eq:CTBham}) is shown in Fig~\ref{fig:CNL_band}(a). One can clearly observe the nodal line on the $\Gamma$-$A$ path. By checking the dispersion around the line, one can verify the leading order band splitting is of cubic order [see Fig.~\ref{fig:CNL_band}(b)].
The cubic character can also be confirmed analytically by expanding the model (\ref{eq:CTBham}) around a generic point $Q$ on the $\Gamma$-$A$ path. The obtained effective model exactly recovers that in Eq.~(\ref{eq:CNLkpham}). 
Here, if we assume the electron filling is one electron per unit cell, i.e., half-filling of the bands, then this MCNL will be staying around the Fermi level.

Besides the MCNL on $\Gamma$-$A$, in Fig.~\ref{fig:CNL_band}(a), one observes there are additional degeneracies between the two bands on the $L$-$M$ and $A$-$H$ paths.
The degeneracies on $L$-$M$ correspond to an essential nodal line. This path has $C_{2v}$ symmetry, which only has a two-dimensional irreducible double representation $\Gamma_5$ \cite{bradley_mathematical_nodate}, so all bands here must acquire a twofold degeneracy. The nodal line on  $L$-$M$ is a conventional linear nodal line. Hence, unlike the MQNL discussed in Sec. \ref{MQNL}, the MCNL must coexist with additional linear nodal lines in the band structure.

As for the degeneracies on $A$-$H$, we find that it actually correspond to a nodal surface on the $k_z=\pi$ plane. This magnetic nodal surface is protected by the $C_{2z}\mathcal{T}'$ symmetry,  Such kind of magnetic nodal surface was first proposed by Wu \emph{et al.} in Ref~\cite{wu_nodal_2018}. Here, we note that ${\cal{T}}^{\prime}$ is an extra symmetry of the lattice model, which is actually not required for the MCNL. Thus, the nodal surface can be removed by breaking the ${\cal{T}}^{\prime}$ symmetry while maintaining $C_{6z}$ and $\tilde{M}_x$ and hence the existence of the MCNL.

\section{Torus surface states}

\begin{figure}[t]
	\includegraphics[width=\columnwidth]{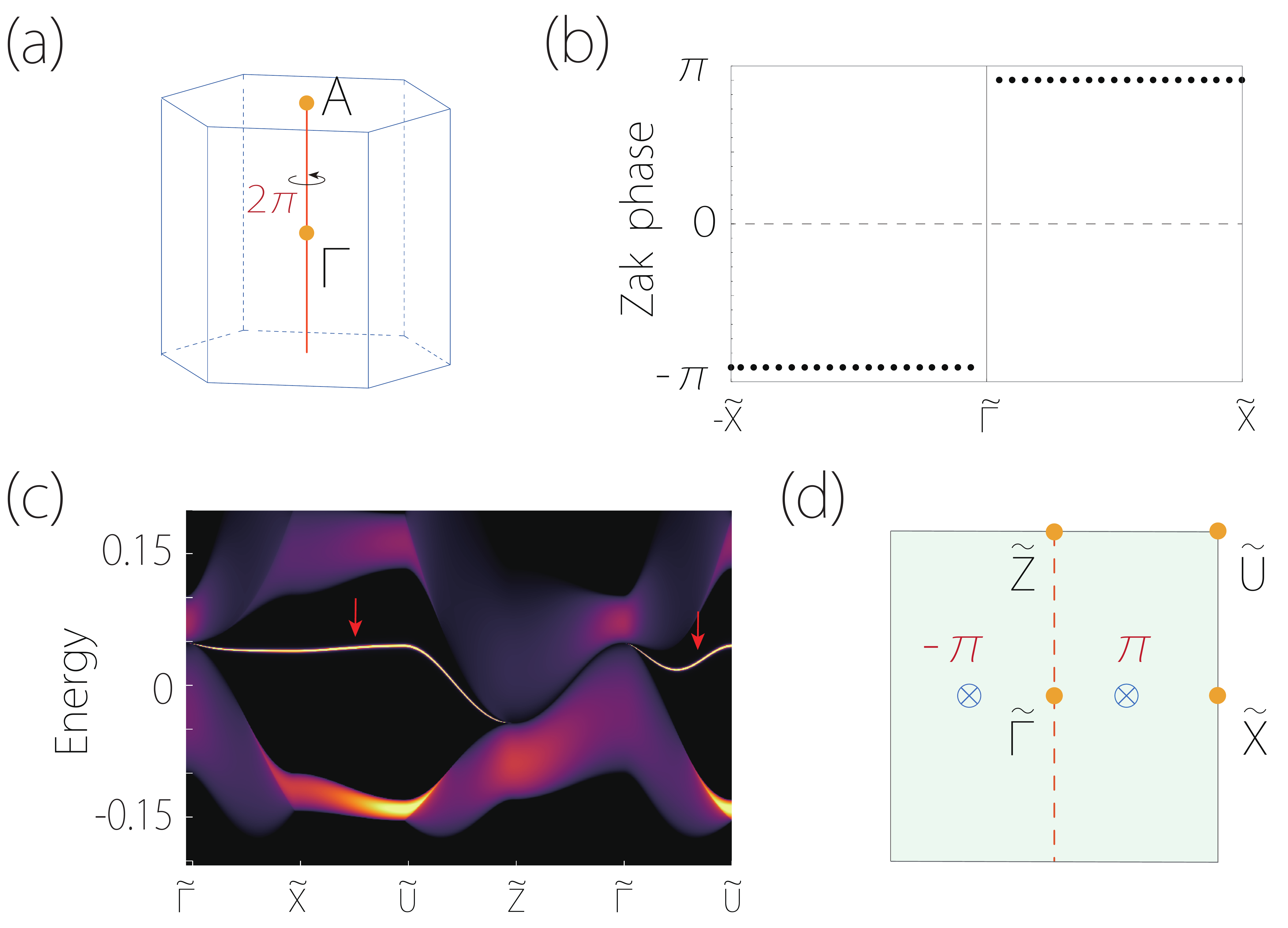}
	\caption{
		(a)  Schematic figure showing the MQNL in the BZ. 	(b) Zak phase ${\cal Z}(k_y,k_z)$ as a function of $k_y$  with $k_z=0$.
		(c) Projected spectrum on the $(10\overline{1}0)$ surface. The red  arrows indicate the torus surface states, which span over the whole surface BZ, as illustrated in (d). In (d), we also indicate the calculated  Zak phase for lines perpendicular to the $(10\overline{1}0)$ surface.	}
	\label{fig:QNL_TSS}
\end{figure}

Materials with conventional linear nodal lines often host drumhead type surface states \cite{yang_dirac_2014,weng_topological_2015}. Here, ``drumhead" means these states occupy a finite region in the surface BZ bounded by the projection of the nodal line on that surface. The stability of these drumhead surface states is often enforced by the nontrivial Zak phase, which is the Berry phase along a straight line crossing the bulk BZ and perpendicular to the specified surface.  For example, to study the $(10\overline{1}0)$ surface normal to the $x$ direction, one may examine the Zak phase \cite{zak_berrys_1989}
\begin{equation}
{\cal Z}(k_y,k_z)=\sum_{n\in\text{occ}}\oint\langle u_{n}(\boldsymbol{k})|i\partial_{k_{x}}|u_{n}(\boldsymbol{k})\rangle dk_{x},
\end{equation}
where $|u\rangle$ is the cell-periodic Bloch state, the integration is over the line with fixed $k_y$ and $k_z$, and the summation is over the occupied bands.

The Zak phase is quantized in units of $\pi$ under certain symmetries. This is the case when the system has both spin rotational symmetry and spacetime inversion symmetry. Another case which is relevant to our discussion here is when the system has a mirror plane normal to the straight line on which the Zak phase is defined. For these cases, an obtained nontrivial Zak phase ${\cal Z}(k_y,k_z)=\pi$ would indicate that there is a surface state at $(k_y,k_z)$ in the BZ for the $(10\overline{1}0)$ surface. And as a linear nodal lines features a $\pi$ Berry phase, it separates regions with $\mathcal{Z}=\pi$ and $\mathcal{Z}=0$ in the surface BZ, and the drumhead surface states reside in the region of $\mathcal{Z}=\pi$.

As shown in Ref.~\cite{yu_quadratic_2019}, the order of the nodal line directly affects its Berry phase. Distinct from the linear nodal lines, the MQNL features a $2\pi$ (equivalent to $0$ when mod $2\pi$) Berry phase for a small loop surrounding it. Consequently, the possible surface states will also exhibit distinct features. In Fig.~\ref{fig:QNL_TSS}(c), we plot the spectrum for the lattice model in Eq.~(\ref{eq:TBham}) on the surface normal to $x$, in which one can clearly observe the surface band. By scanning the surface BZ, we find that this surface band covers the whole BZ. Since the surface BZ forms a torus $T^2$, these states may be termed as the torus surface states. The presence of torus surface states is consistent with the calculated Zak phase as shown in Fig.~\ref{fig:QNL_TSS}(b) and \ref{fig:QNL_TSS}(d). Here, the red dashed line in Fig.~\ref{fig:QNL_TSS}(d)  indicates the projection of the MQNL in the surface BZ. The important point is that unlike the linear nodal line, the MQNL does not impose a $\pi$ discontinuity in the Zak phase, therefore $\mathcal{Z}$ can take the nontrivial value $\pi$ in the whole surface BZ, leading to the torus surface states.

\begin{figure}[t]
	\includegraphics[width=\columnwidth]{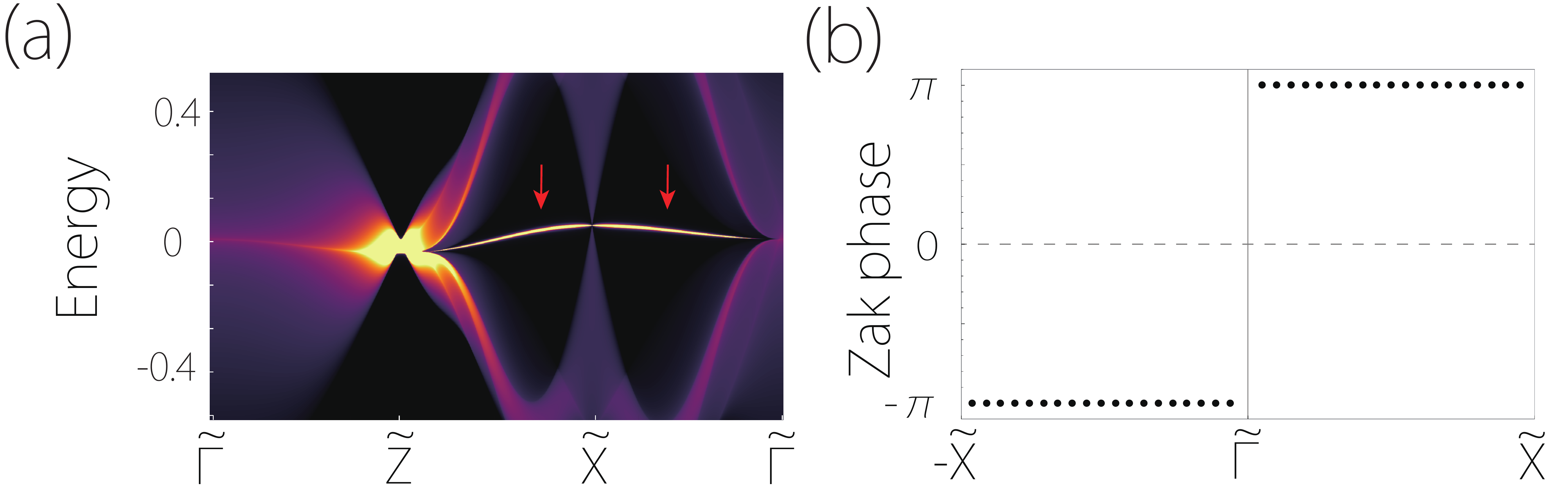}
	\caption{(a) Projected spectrum on the $(10\overline{1}0)$ surface for the MCNL model. The red  arrows indicate the torus surface states. (b) Zak phase ${\cal Z}(k_y,k_z)$ as a function of $k_y$  with $k_z=0$.	}
	\label{fig:CNL_TSS}
\end{figure}

\begin{figure*}[t]
	\includegraphics[width=16 cm]{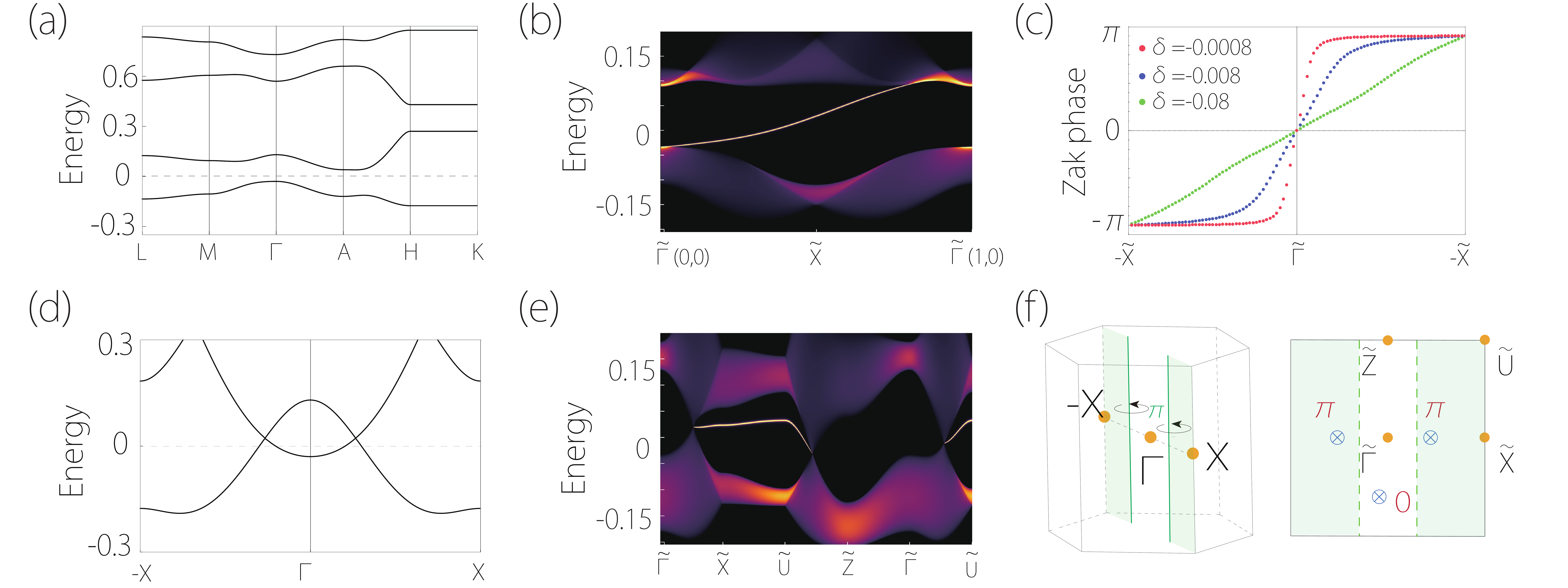}
	\caption{MQNL under symmetry breaking.
(a-c) Breaking $M_{z}{\cal{T}}$ transforms the MQNL  into a  3D QAH insulator. (a) Band structure of the MQNL model (\ref{eq:TBham}) with  perturbation (\ref{eq:QNLpert1}). (b) shows the projected spectrum on the $(10\overline{1}0)$ surface. In the calculation, we set $\delta=-0.08$.   (c) The Wilson loop in the $k_z=0$ plane for different perturbation strength $\delta$, showing a unit Chern number.
	(d-f) Breaking $C_{3z}$ transforms the MQNL  into two linear nodal lines. (d) Band structure [along the path marked in (f)] and (e) the projected spectrum on the $(10\overline{1}0)$ surface of MQNL model (\ref{eq:TBham}) with  perturbation (\ref{eq:QNLpert2}).  (f)  Schematic figure showing the two linear nodal lines (green lines) and the corresponding surface state distribution the surface BZ (green colored region). }
	\label{fig:QNL_break}
\end{figure*}

Similar analysis can be performed for the lattice model (\ref{eq:CTBham}) containing the MCNL. MCNL carries a $\pi$ Berry phase, so one may expect to see drumhead surface states similar to that of linear nodal lines. In Fig.~\ref{fig:CNL_TSS}(a), we show the result of the side surface normal to $x$. Interestingly, again, one finds that there exist torus surface states, and the Zak phase is nontrivial in the whole surface BZ. The reason is that as shown in Sec.~\ref{LatticeModelForMCNL}, besides the MCNL, the system must also have essential linear nodal lines along the $L$-$M$ path. On the surface in Fig.~\ref{fig:CNL_BZ}(c), the MCNL and the linear nodal line project to the same line $\widetilde{\Gamma}$-$\widetilde{Z}$ in the surface BZ. Hence, the discontinuity in the Zak phase across $\widetilde{\Gamma}$-$\widetilde{Z}$ becomes $\pi+\pi=0\ \text{mod}\ 2\pi$ [see Fig.~\ref{fig:CNL_TSS}(b)], similar to that in Fig.~\ref{fig:QNL_TSS}(d). Therefore, torus surface states can appear in this MCNL system as well.

We have shown that torus surface states are compatible with both MQNLs and MCNLs, and their existence is indeed demonstrated in the two lattice models. However, we must point out that MQNLs and MCNLs cannot guarantee the existence of torus surface states. It is possible to have the Zak phase tuned to zero for the whole surface BZ, then there will be no stable surface states for the system.

\section{Topological phase transition }\label{TPT}

The magnetic higher-order nodal lines discovered here are protected by multiple symmetries. Under symmetry breaking, they may transform to other interesting band features. Particularly, the lack of time reversal symmetry may generate new physics not possible in their nonmagnetic counterparts.

Let's first consider the MQNL. When we break the $M_{z}{\cal T}$ symmetry, the degeneracy of the MQNL will be lifted, and the system may be turned into an insulator. For example, in the model Eq.~(\ref{eq:TBham}), we may break the $M_{z}{\cal T}$ symmetry by adding the following perturbation term:
\begin{equation}
{\cal H}_{1}=\delta\ \Gamma_{33}, \label{eq:QNLpert1}
\end{equation}
where $\delta$ denotes the perturbation strength. The resulting band structure is plotted in Fig.~\ref{fig:QNL_break}(a). One observes that the original MQNL is fully gapped out and the system becomes an insulator. Interestingly, we find that this insulator is topologically nontrivial.
In Fig.~\ref{fig:QNL_break}(c), we evaluate the Chern number ${\cal C}$ of the $k_z=0$ plane by using the Wilson loop method. (Here, we have only the lowest band occupied.)
The result shows that $\mathcal{C}=1$ is nontrivial. Correspondingly, there must exist chiral edge state for this 2D subsystem, as verified in Fig.~\ref{fig:QNL_break}(b).
Now, since the system has a global band gap, every 2D slice of the bulk BZ with a fixed $k_z$ must have the same Chern number $\mathcal{C}=1$. Thus, the side surface of the system must be covered by chiral boundary states propagating in the same direction.
This leads to a novel 3D QAH insulator state, which features a vanishing longitudinal conductivity $\sigma_{xx}=0$ and an quantized anomalous Hall conductivity given by
\begin{equation}
  \sigma_{xy}=\int_{-\pi/d}^{\pi/d}\frac{dk_z}{2\pi}\left(\mathcal{C}\frac{e^2}{h}\right)=\mathcal{C}\frac{e^2}{hd},
\end{equation}
where we have explicitly written out the factor of $1/d$ in the wave vector, with $d$ the lattice constant along $z$. Thus, the quantization of $\sigma_{xy}$ for the 3D QAH state is in units of $e^2/(hd)$.

We note that this 3D QAH state is analogous to the 3D quantum Hall effect (QHE) originally proposed by Halperin~\cite{Halperin_1987} and recently demonstrated by Tang \emph{et al.}~\cite{TangNatureHall_2019}. Meanwhile, there are also important differences. The 3D QHE is realized under strong magnetic field with Landau band formation, whereas the 3D QAH state exists without external magnetic field. Related to this point, for the 3D QHE, the quantization unit (and hence $\sigma_{xy}$) typically changes with the magnetic field strength, since the Landau band degeneracy scales linearly with $B$~\cite{Halperin_1987}. In comparison, for a 3D QAH insulator, the quantization unit is a  fixed value, determined by the structure of the system.

Another interesting case is to break the $C_{3z}$ symmetry in model, while keeping the $M_{x}$ symmetry. A possible perturbation term may be written as
\begin{equation}\label{eq:QNLpert2}
{\cal H}_{2}=\delta\left(\cos \frac{k_z}{2}\Gamma_{10} + \sin \frac{k_z}{2}\Gamma_{20}\right).
\end{equation}
With this term, the MQNL will split into two linear nodal lines lying in the $M_x$ mirror plane, as illustrated in Fig.~\ref{fig:QNL_break}(d). On the surface that is parallel to the mirror plane, the original torus surface states will transform into drumhead surface states [see Fig.~\ref{fig:QNL_break}(e,f)]. Compared to Fig.~\ref{fig:QNL_TSS}(a), the splitting of MQNL opens a region between the two linear nodal lines which has a trivial Zak phase. As a result, surface states now exist only in the region outside of the two nodal lines in the surface BZ.

\begin{figure}[t]
	\includegraphics[width=\columnwidth]{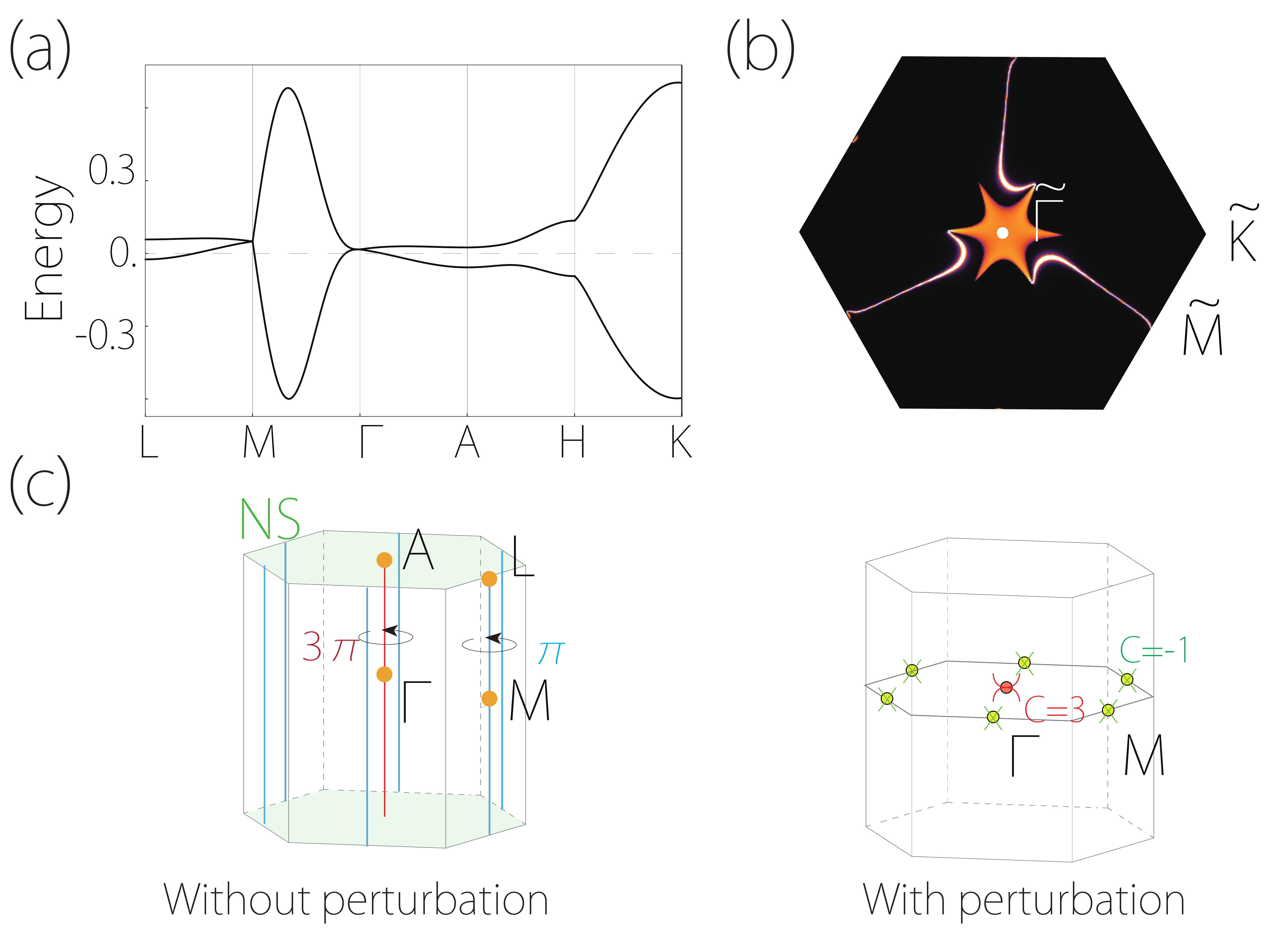}
	\caption{ MCNL semimetal evolves into a triple-Weyl semimetal under symmetry breaking.	
(a) Band structure of MCNL model (\ref{eq:CTBham}) with  perturbation (\ref{per3}). (b) shows the projected spectrum on the $(0001)$  surface, showing three Fermi arcs connecting the projections of the triple Weyl point and three single Weyl points. In the calculation, we set $\delta_1=\delta_2=0.05$.   (c) Schematic figure showing the transition from the MCNL semimetal to the triple-Weyl semimetal. 	}
	\label{fig:CNL_break}
\end{figure}

As for the MCNL in model, let us consider the case when both $C_{6z}$ and $\tilde{M}_x$ symmetries are broken but $C_{3z}$ and ${\cal{T}}^{\prime}$ symmetries are maintained. This  allows the following perturbation term
\begin{equation}\label{per3}
{\cal H}_{3}=\delta _1  \sin \frac{k_z}{2} \sigma_1-2 \delta _2\sin \frac{k_y}{2} \left( \cos \frac{\sqrt{3} k_x}{2} -  \cos \frac{k_y}{2}\right)\sigma _3.
\end{equation}
In this case, the MCNL, the essential linear nodal lines, and the magnetic nodal surface are all destroyed. The MCNL and the linear nodal lines respectively evolve into a triple-Weyl point with Chern number ${\cal{C}}=3$ and three conventional Weyl points with  ${\cal{C}}=-1$ (see Fig.~\ref{fig:CNL_break}). These Weyl points are pinned at the time reversal invariant momenta in the $k_z=0$ plane, namely, the $\Gamma$ and the three $M$ points, which is  due to the Kramers-like degeneracy for ${\cal{T}}^{\prime 2}=-1$ in the $k_z=0$ plane. In comparison, ${\cal{T}}^{\prime 2}=1$ in the $k_z=\pi$ plane, so there is no such degeneracy there. We note that this configuration with a single protected triple-Weyl point in the BZ is unique for magnetic systems; for nonmagnetic system, the protected triple Weyl point would appear at least in a pair, {or comes with a nodal surface}  \cite{yu_quadratic_2019}.

\section{Discussion and Conclusion}

In this work, we have theoretically demonstrated the existence of symmetry protected MQNLs and MCNLs. It remains an important task to identify realistic materials that can host these higher-order nodal lines. The symmetry conditions found here will offer useful guidance for such material search. Besides the nodal lines, it is also interesting to search for the topological states identified in Sec. \ref{TPT}, such as the 3D QAH state and the magnetic triple-Weyl point.

In experiment, the order of dispersion and the surface states can be directly probed by the angle-resolved photoemission spectroscopy (ARPES) method \cite{HongNRP_2019}. As discussed in Ref.~\cite{yu_quadratic_2019}, the higher-order nodal lines can give different scalings in the joint density of states (DOS) and Landau level energies, which can be probed with infrared optical spectroscopy. The torus surface states can lead to a large surface DOS, which can be probed by the scanning tunneling spectroscopy (STS) \cite{Zheng_ACS_2016}. Such large surface DOS could be beneficial for realizing surface magnetism and surface high-temperature superconductivity.

{In conclusion, with  symmetry analysis, we have proposed a new class of nodal lines,  namely, the MQNLs and MCNLs in magnetic systems. We present sufficient symmetry conditions for stabilizing these high-order nodal lines, and construct concrete lattice models to further demonstrate our proposals. Remarkably, we  find  the MQNL can exist as the only band degeneracy at the Fermi level. Both MQNL and MCNL can give rise to novel torus surface state, covering the whole surface BZ. Furthermore, these nodal lines may be regarded as parent phases for many other interesting topological phases. Under symmetry breaking, they can generate phases with 3D QAH insulator state,  multiple magnetic linear nodal lines, and magnetic triple-Weyl points.
Our work extends the scope of higher-order nodal lines to magnetic systems,  reveals a route towards 3D QAH insulator state, and provides useful guidance to search for magnetic topological materials.}

\acknowledgments
The authors thank K. Fredericks and D. L. Deng for valuable discussions. This work is supported by China Postdoctoral Science Foundation (Grant No. 2020M670106), the NSF of China (Grant Nos. 12004028, 12004035), Fundamental Research Funds for the Central Universities (ZY2018), Beijing Institute of Technology Research Fund Program for Young Scholars, and Singapore Ministry of Education AcRF Tier 2 (Grant No. MOE2019-T2-1-001).

\appendix
\section{Two-band effective Model of MQNL } \label{appendix1}

Since the magnetic nodal line formed by the lowest two bands  in Fig.~\ref{fig:QNL_band}(a) are well separated from the other  bands in energy,
we can establish a two-band effective model to capture the physics of the line.
First, we expand  the  lattice model (\ref{eq:TBham}) around a point on the $\Gamma$-$A$ path (up to quadratic order)
\begin{eqnarray}\label{QGATB}
{\cal H}_{\Gamma A} &= & A_0 +\frac{A_1}{2}(k_x \Gamma _{0,2}- k_y \Gamma _{31})\nonumber\\
&& +\frac{A_1}{8 \sqrt{3}}\left[\left(k_x^2-k_y^2\right)  \Gamma_{01}-2k_x k_y \Gamma_{32}\right] \nonumber\\
&& +3 A_2 \cos \frac{k_z}{2} \Gamma _{11}+3 A_3 \sin \frac{k_z}{2} \Gamma_{2,1}  \nonumber\\
&&- k_{\|}^{2} (\frac{A_2}{4} \cos \frac{k_z}{2} \Gamma _{11}+\frac{A_3}{4}\sin \frac{k_z}{2} \Gamma_{21}),
\end{eqnarray}
with $k_{\|}=\sqrt{k_x^2+k_y^2}$.
From Eq. (\ref{QGATB}), the eigenstates of the lowest two bands  are found to be
\begin{eqnarray}
\Psi_1={\cal N }^{-1}(\gamma, 0, 0, 1)^T, \ \ \ \Psi_2={\cal N }^{-1}(0, \gamma, 1, 0)^T,
\end{eqnarray}
where 
\begin{equation}
  \gamma=-\frac{\sqrt{A_{2}^2+(A_{2}^2-A_{3}^2)  \cos k_z+A_{3}^2}}{\sqrt{2} \left(A_{2} \cos \frac{k_z}{2}+i A_{3} \sin \frac{k_z}{2}\right)},
\end{equation}
and ${\cal N }$ is a  normalization coefficient. Then the two-band effective model can be obtained as
\begin{eqnarray}
{\cal H}_{\text{eff}}(\boldsymbol{k})&=&
\left[\begin{array}{cc}
\langle \Psi_{1}| {\cal H}_{\Gamma A}(\boldsymbol{k}) | \Psi_{1}\rangle &  \langle \Psi_{1}| {\cal H}_{\Gamma A}(\boldsymbol{k}) | \Psi_{2}\rangle \\
\langle \Psi_{2}| {\cal H}_{\Gamma A}(\boldsymbol{k}) | \Psi_{1}\rangle & \langle \Psi_{2}| {\cal H}_{\Gamma A}(\boldsymbol{k}) | \Psi_{2}\rangle
\end{array}\right]\nonumber\\
&=&c k_{\|}^2 + A(k_{+}^2\sigma_{-}+k_{-}^2\sigma_{+}),
\label{eq:effham}
\end{eqnarray}
which recovers the effective Hamiltonian of MQNL in the Eq.~(\ref{eq:kpham}). In this effective model, the parameters are found to be  $c=\sqrt{A_{2}^2+(A_{2}^2-A_{3}^2) \cos k_z+A_{3}^2}/(2 \sqrt{2} |{\cal{N}}|^2)$ and $A=A_1/(4 \sqrt{3}|{\cal{N}}|^2)$.


\bibliography{mqcnlv2}

\begin{thebibliography}{62}%
\makeatletter
\providecommand \@ifxundefined [1]{%
 \@ifx{#1\undefined}
}%
\providecommand \@ifnum [1]{%
 \ifnum #1\expandafter \@firstoftwo
 \else \expandafter \@secondoftwo
 \fi
}%
\providecommand \@ifx [1]{%
 \ifx #1\expandafter \@firstoftwo
 \else \expandafter \@secondoftwo
 \fi
}%
\providecommand \natexlab [1]{#1}%
\providecommand \enquote  [1]{``#1''}%
\providecommand \bibnamefont  [1]{#1}%
\providecommand \bibfnamefont [1]{#1}%
\providecommand \citenamefont [1]{#1}%
\providecommand \href@noop [0]{\@secondoftwo}%
\providecommand \href [0]{\begingroup \@sanitize@url \@href}%
\providecommand \@href[1]{\@@startlink{#1}\@@href}%
\providecommand \@@href[1]{\endgroup#1\@@endlink}%
\providecommand \@sanitize@url [0]{\catcode `\\12\catcode `\$12\catcode
  `\&12\catcode `\#12\catcode `\^12\catcode `\_12\catcode `\%12\relax}%
\providecommand \@@startlink[1]{}%
\providecommand \@@endlink[0]{}%
\providecommand \url  [0]{\begingroup\@sanitize@url \@url }%
\providecommand \@url [1]{\endgroup\@href {#1}{\urlprefix }}%
\providecommand \urlprefix  [0]{URL }%
\providecommand \Eprint [0]{\href }%
\providecommand \doibase [0]{https://doi.org/}%
\providecommand \selectlanguage [0]{\@gobble}%
\providecommand \bibinfo  [0]{\@secondoftwo}%
\providecommand \bibfield  [0]{\@secondoftwo}%
\providecommand \translation [1]{[#1]}%
\providecommand \BibitemOpen [0]{}%
\providecommand \bibitemStop [0]{}%
\providecommand \bibitemNoStop [0]{.\EOS\space}%
\providecommand \EOS [0]{\spacefactor3000\relax}%
\providecommand \BibitemShut  [1]{\csname bibitem#1\endcsname}%
\let\auto@bib@innerbib\@empty
\bibitem [{\citenamefont {Chiu}\ \emph {et~al.}(2016)\citenamefont {Chiu},
  \citenamefont {Teo}, \citenamefont {Schnyder},\ and\ \citenamefont
  {Ryu}}]{chiu_classification_2016}%
  \BibitemOpen
  \bibfield  {author} {\bibinfo {author} {\bibfnamefont {C.-K.}\ \bibnamefont
  {Chiu}}, \bibinfo {author} {\bibfnamefont {J.~C.~Y.}\ \bibnamefont {Teo}},
  \bibinfo {author} {\bibfnamefont {A.~P.}\ \bibnamefont {Schnyder}},\ and\
  \bibinfo {author} {\bibfnamefont {S.}~\bibnamefont {Ryu}},\ }\bibfield
  {title} {\bibinfo {title} {Classification of topological quantum matter with
  symmetries},\ }\href {https://doi.org/10.1103/RevModPhys.88.035005}
  {\bibfield  {journal} {\bibinfo  {journal} {Rev. Mod. Phys.}\ }\textbf
  {\bibinfo {volume} {88}},\ \bibinfo {pages} {035005} (\bibinfo {year}
  {2016})}\BibitemShut {NoStop}%
\bibitem [{\citenamefont {Armitage}\ \emph {et~al.}(2018)\citenamefont
  {Armitage}, \citenamefont {Mele},\ and\ \citenamefont
  {Vishwanath}}]{armitage_weyl_2018}%
  \BibitemOpen
  \bibfield  {author} {\bibinfo {author} {\bibfnamefont {N.~P.}\ \bibnamefont
  {Armitage}}, \bibinfo {author} {\bibfnamefont {E.~J.}\ \bibnamefont {Mele}},\
  and\ \bibinfo {author} {\bibfnamefont {A.}~\bibnamefont {Vishwanath}},\
  }\bibfield  {title} {\bibinfo {title} {Weyl and {Dirac} semimetals in
  three-dimensional solids},\ }\href
  {https://doi.org/10.1103/RevModPhys.90.015001} {\bibfield  {journal}
  {\bibinfo  {journal} {Rev. Mod. Phys.}\ }\textbf {\bibinfo {volume} {90}},\
  \bibinfo {pages} {015001} (\bibinfo {year} {2018})}\BibitemShut {NoStop}%
\bibitem [{\citenamefont {Yang}(2016)}]{yang_dirac_2016}%
  \BibitemOpen
  \bibfield  {author} {\bibinfo {author} {\bibfnamefont {S.~A.}\ \bibnamefont
  {Yang}},\ }\bibfield  {title} {\bibinfo {title} {Dirac and {Weyl}
  {Materials}: {Fundamental} {Aspects} and {Some} {Spintronics}
  {Applications}},\ }\href {https://doi.org/10.1142/S2010324716400038}
  {\bibfield  {journal} {\bibinfo  {journal} {SPIN}\ }\textbf {\bibinfo
  {volume} {06}},\ \bibinfo {pages} {1640003} (\bibinfo {year}
  {2016})}\BibitemShut {NoStop}%
\bibitem [{\citenamefont {Burkov}(2016)}]{Burkov2016Topological-Nm}%
  \BibitemOpen
  \bibfield  {author} {\bibinfo {author} {\bibfnamefont {A.}~\bibnamefont
  {Burkov}},\ }\bibfield  {title} {\bibinfo {title} {Topological semimetals},\
  }\href {https://doi.org/10.1038/nmat4788} {\bibfield  {journal} {\bibinfo
  {journal} {Nat. Mater.}\ }\textbf {\bibinfo {volume} {15}},\ \bibinfo {pages}
  {1145} (\bibinfo {year} {2016})}\BibitemShut {NoStop}%
\bibitem [{\citenamefont {Nielsen}\ and\ \citenamefont
  {Ninomiya}(1983)}]{Nielsen1983Adler-PLB}%
  \BibitemOpen
  \bibfield  {author} {\bibinfo {author} {\bibfnamefont {H.~B.}\ \bibnamefont
  {Nielsen}}\ and\ \bibinfo {author} {\bibfnamefont {M.}~\bibnamefont
  {Ninomiya}},\ }\bibfield  {title} {\bibinfo {title} {The adler-bell-jackiw
  anomaly and weyl fermions in a crystal},\ }\href
  {http://www.sciencedirect.com/science/article/pii/0370269383915290}
  {\bibfield  {journal} {\bibinfo  {journal} {Phys. Lett. B}\ }\textbf
  {\bibinfo {volume} {130}},\ \bibinfo {pages} {389} (\bibinfo {year}
  {1983})}\BibitemShut {NoStop}%
\bibitem [{\citenamefont {Volovik}(2003)}]{Volovik2003universe-OUPoD}%
  \BibitemOpen
  \bibfield  {author} {\bibinfo {author} {\bibfnamefont {G.~E.}\ \bibnamefont
  {Volovik}},\ }\href@noop {} {\emph {\bibinfo {title} {The Universe in a
  Helium Droplet}}}\ (\bibinfo  {publisher} {Clarendon Press},\ \bibinfo
  {address} {Oxford},\ \bibinfo {year} {2003})\BibitemShut {NoStop}%
\bibitem [{\citenamefont {Hosur}\ \emph {et~al.}(2012)\citenamefont {Hosur},
  \citenamefont {Parameswaran},\ and\ \citenamefont
  {Vishwanath}}]{HosurPRB-2012}%
  \BibitemOpen
  \bibfield  {author} {\bibinfo {author} {\bibfnamefont {P.}~\bibnamefont
  {Hosur}}, \bibinfo {author} {\bibfnamefont {S.~A.}\ \bibnamefont
  {Parameswaran}},\ and\ \bibinfo {author} {\bibfnamefont {A.}~\bibnamefont
  {Vishwanath}},\ }\bibfield  {title} {\bibinfo {title} {Charge transport in
  weyl semimetals},\ }\href {https://doi.org/10.1103/PhysRevLett.108.046602}
  {\bibfield  {journal} {\bibinfo  {journal} {Phys. Rev. Lett.}\ }\textbf
  {\bibinfo {volume} {108}},\ \bibinfo {pages} {046602} (\bibinfo {year}
  {2012})}\BibitemShut {NoStop}%
\bibitem [{\citenamefont {Son}\ and\ \citenamefont
  {Spivak}(2013)}]{Son2013Chiral-PRB}%
  \BibitemOpen
  \bibfield  {author} {\bibinfo {author} {\bibfnamefont {D.}~\bibnamefont
  {Son}}\ and\ \bibinfo {author} {\bibfnamefont {B.}~\bibnamefont {Spivak}},\
  }\bibfield  {title} {\bibinfo {title} {Chiral anomaly and classical negative
  magnetoresistance of {Weyl} metals},\ }\href
  {https://link.aps.org/doi/10.1103/PhysRevB.88.104412} {\bibfield  {journal}
  {\bibinfo  {journal} {Phys. Rev. B}\ }\textbf {\bibinfo {volume} {88}},\
  \bibinfo {pages} {104412} (\bibinfo {year} {2013})}\BibitemShut {NoStop}%
\bibitem [{\citenamefont {de~Juan}\ \emph {et~al.}(2017)\citenamefont
  {de~Juan}, \citenamefont {Grushin}, \citenamefont {Morimoto},\ and\
  \citenamefont {Moore}}]{Quantized_NC_2017}%
  \BibitemOpen
  \bibfield  {author} {\bibinfo {author} {\bibfnamefont {F.}~\bibnamefont
  {de~Juan}}, \bibinfo {author} {\bibfnamefont {A.~G.}\ \bibnamefont
  {Grushin}}, \bibinfo {author} {\bibfnamefont {T.}~\bibnamefont {Morimoto}},\
  and\ \bibinfo {author} {\bibfnamefont {J.~E.}\ \bibnamefont {Moore}},\
  }\bibfield  {title} {\bibinfo {title} {Quantized circular photogalvanic
  effect in {Weyl} semimetals},\ }\href {https://doi.org/10.1038/ncomms15995}
  {\bibfield  {journal} {\bibinfo  {journal} {Nat. Commun.}\ }\textbf {\bibinfo
  {volume} {8}},\ \bibinfo {pages} {15995} (\bibinfo {year}
  {2017})}\BibitemShut {NoStop}%
\bibitem [{\citenamefont {Liu}\ \emph {et~al.}(2020)\citenamefont {Liu},
  \citenamefont {Yu}, \citenamefont {Xiao},\ and\ \citenamefont
  {Yang}}]{LiuY-PRL2020}%
  \BibitemOpen
  \bibfield  {author} {\bibinfo {author} {\bibfnamefont {Y.}~\bibnamefont
  {Liu}}, \bibinfo {author} {\bibfnamefont {Z.-M.}\ \bibnamefont {Yu}},
  \bibinfo {author} {\bibfnamefont {C.}~\bibnamefont {Xiao}},\ and\ \bibinfo
  {author} {\bibfnamefont {S.~A.}\ \bibnamefont {Yang}},\ }\bibfield  {title}
  {\bibinfo {title} {Quantized circulation of anomalous shift in interface
  reflection},\ }\href {https://doi.org/10.1103/PhysRevLett.125.076801}
  {\bibfield  {journal} {\bibinfo  {journal} {Phys. Rev. Lett.}\ }\textbf
  {\bibinfo {volume} {125}},\ \bibinfo {pages} {076801} (\bibinfo {year}
  {2020})}\BibitemShut {NoStop}%
\bibitem [{\citenamefont {Wan}\ \emph {et~al.}(2011)\citenamefont {Wan},
  \citenamefont {Turner}, \citenamefont {Vishwanath},\ and\ \citenamefont
  {Savrasov}}]{wan_topological_2011}%
  \BibitemOpen
  \bibfield  {author} {\bibinfo {author} {\bibfnamefont {X.}~\bibnamefont
  {Wan}}, \bibinfo {author} {\bibfnamefont {A.~M.}\ \bibnamefont {Turner}},
  \bibinfo {author} {\bibfnamefont {A.}~\bibnamefont {Vishwanath}},\ and\
  \bibinfo {author} {\bibfnamefont {S.~Y.}\ \bibnamefont {Savrasov}},\
  }\bibfield  {title} {\bibinfo {title} {Topological semimetal and {Fermi}-arc
  surface states in the electronic structure of pyrochlore iridates},\ }\href
  {https://doi.org/10.1103/PhysRevB.83.205101} {\bibfield  {journal} {\bibinfo
  {journal} {Phys. Rev. B}\ }\textbf {\bibinfo {volume} {83}},\ \bibinfo
  {pages} {205101} (\bibinfo {year} {2011})}\BibitemShut {NoStop}%
\bibitem [{\citenamefont {Young}\ \emph {et~al.}(2012)\citenamefont {Young},
  \citenamefont {Zaheer}, \citenamefont {Teo}, \citenamefont {Kane},
  \citenamefont {Mele},\ and\ \citenamefont {Rappe}}]{Young2012Dirac-Prl}%
  \BibitemOpen
  \bibfield  {author} {\bibinfo {author} {\bibfnamefont {S.~M.}\ \bibnamefont
  {Young}}, \bibinfo {author} {\bibfnamefont {S.}~\bibnamefont {Zaheer}},
  \bibinfo {author} {\bibfnamefont {J.~C.}\ \bibnamefont {Teo}}, \bibinfo
  {author} {\bibfnamefont {C.~L.}\ \bibnamefont {Kane}}, \bibinfo {author}
  {\bibfnamefont {E.~J.}\ \bibnamefont {Mele}},\ and\ \bibinfo {author}
  {\bibfnamefont {A.~M.}\ \bibnamefont {Rappe}},\ }\bibfield  {title} {\bibinfo
  {title} {Dirac semimetal in three dimensions},\ }\href
  {https://journals.aps.org/prl/abstract/10.1103/PhysRevLett.108.140405}
  {\bibfield  {journal} {\bibinfo  {journal} {Phys. Rev. Lett.}\ }\textbf
  {\bibinfo {volume} {108}},\ \bibinfo {pages} {140405} (\bibinfo {year}
  {2012})}\BibitemShut {NoStop}%
\bibitem [{\citenamefont {Wang}\ \emph {et~al.}(2012)\citenamefont {Wang},
  \citenamefont {Sun}, \citenamefont {Chen}, \citenamefont {Franchini},
  \citenamefont {Xu}, \citenamefont {Weng}, \citenamefont {Dai},\ and\
  \citenamefont {Fang}}]{wang_dirac_2012}%
  \BibitemOpen
  \bibfield  {author} {\bibinfo {author} {\bibfnamefont {Z.}~\bibnamefont
  {Wang}}, \bibinfo {author} {\bibfnamefont {Y.}~\bibnamefont {Sun}}, \bibinfo
  {author} {\bibfnamefont {X.-Q.}\ \bibnamefont {Chen}}, \bibinfo {author}
  {\bibfnamefont {C.}~\bibnamefont {Franchini}}, \bibinfo {author}
  {\bibfnamefont {G.}~\bibnamefont {Xu}}, \bibinfo {author} {\bibfnamefont
  {H.}~\bibnamefont {Weng}}, \bibinfo {author} {\bibfnamefont {X.}~\bibnamefont
  {Dai}},\ and\ \bibinfo {author} {\bibfnamefont {Z.}~\bibnamefont {Fang}},\
  }\bibfield  {title} {\bibinfo {title} {Dirac semimetal and topological phase
  transitions in {A}$_{\textrm{3}}${Bi} ({A}={Na}, {K}, {Rb})},\ }\href
  {https://doi.org/10.1103/PhysRevB.85.195320} {\bibfield  {journal} {\bibinfo
  {journal} {Phys. Rev. B}\ }\textbf {\bibinfo {volume} {85}},\ \bibinfo
  {pages} {195320} (\bibinfo {year} {2012})}\BibitemShut {NoStop}%
\bibitem [{\citenamefont {Zhao}\ and\ \citenamefont {Wang}(2013)}]{Zhao2013}%
  \BibitemOpen
  \bibfield  {author} {\bibinfo {author} {\bibfnamefont {Y.~X.}\ \bibnamefont
  {Zhao}}\ and\ \bibinfo {author} {\bibfnamefont {Z.~D.}\ \bibnamefont
  {Wang}},\ }\bibfield  {title} {\bibinfo {title} {Topological classification
  and stability of fermi surfaces},\ }\href
  {https://doi.org/10.1103/PhysRevLett.110.240404} {\bibfield  {journal}
  {\bibinfo  {journal} {Phys. Rev. Lett.}\ }\textbf {\bibinfo {volume} {110}},\
  \bibinfo {pages} {240404} (\bibinfo {year} {2013})}\BibitemShut {NoStop}%
\bibitem [{\citenamefont {Bradlyn}\ \emph {et~al.}(2016)\citenamefont
  {Bradlyn}, \citenamefont {Cano}, \citenamefont {Wang}, \citenamefont
  {Vergniory}, \citenamefont {Felser}, \citenamefont {Cava},\ and\
  \citenamefont {Bernevig}}]{bradlyn_beyond_2016}%
  \BibitemOpen
  \bibfield  {author} {\bibinfo {author} {\bibfnamefont {B.}~\bibnamefont
  {Bradlyn}}, \bibinfo {author} {\bibfnamefont {J.}~\bibnamefont {Cano}},
  \bibinfo {author} {\bibfnamefont {Z.}~\bibnamefont {Wang}}, \bibinfo {author}
  {\bibfnamefont {M.~G.}\ \bibnamefont {Vergniory}}, \bibinfo {author}
  {\bibfnamefont {C.}~\bibnamefont {Felser}}, \bibinfo {author} {\bibfnamefont
  {R.~J.}\ \bibnamefont {Cava}},\ and\ \bibinfo {author} {\bibfnamefont
  {B.~A.}\ \bibnamefont {Bernevig}},\ }\bibfield  {title} {\bibinfo {title}
  {Beyond {Dirac} and {Weyl} fermions: {Unconventional} quasiparticles in
  conventional crystals},\ }\href {https://doi.org/10.1126/science.aaf5037}
  {\bibfield  {journal} {\bibinfo  {journal} {Science}\ }\textbf {\bibinfo
  {volume} {353}},\ \bibinfo {pages} {aaf5037} (\bibinfo {year}
  {2016})}\BibitemShut {NoStop}%
\bibitem [{\citenamefont {Burkov}\ \emph {et~al.}(2011)\citenamefont {Burkov},
  \citenamefont {Hook},\ and\ \citenamefont
  {Balents}}]{burkov_topological_2011}%
  \BibitemOpen
  \bibfield  {author} {\bibinfo {author} {\bibfnamefont {A.~A.}\ \bibnamefont
  {Burkov}}, \bibinfo {author} {\bibfnamefont {M.~D.}\ \bibnamefont {Hook}},\
  and\ \bibinfo {author} {\bibfnamefont {L.}~\bibnamefont {Balents}},\
  }\bibfield  {title} {\bibinfo {title} {Topological nodal semimetals},\ }\href
  {https://doi.org/10.1103/PhysRevB.84.235126} {\bibfield  {journal} {\bibinfo
  {journal} {Phys. Rev. B}\ }\textbf {\bibinfo {volume} {84}},\ \bibinfo
  {pages} {235126} (\bibinfo {year} {2011})}\BibitemShut {NoStop}%
\bibitem [{\citenamefont {Yang}\ \emph {et~al.}(2014)\citenamefont {Yang},
  \citenamefont {Pan},\ and\ \citenamefont {Zhang}}]{yang_dirac_2014}%
  \BibitemOpen
  \bibfield  {author} {\bibinfo {author} {\bibfnamefont {S.~A.}\ \bibnamefont
  {Yang}}, \bibinfo {author} {\bibfnamefont {H.}~\bibnamefont {Pan}},\ and\
  \bibinfo {author} {\bibfnamefont {F.}~\bibnamefont {Zhang}},\ }\bibfield
  {title} {\bibinfo {title} {Dirac and {Weyl} {Superconductors} in {Three}
  {Dimensions}},\ }\href {https://doi.org/10.1103/PhysRevLett.113.046401}
  {\bibfield  {journal} {\bibinfo  {journal} {Phys. Rev. Lett.}\ }\textbf
  {\bibinfo {volume} {113}},\ \bibinfo {pages} {046401} (\bibinfo {year}
  {2014})}\BibitemShut {NoStop}%
\bibitem [{\citenamefont {Weng}\ \emph {et~al.}(2015)\citenamefont {Weng},
  \citenamefont {Liang}, \citenamefont {Xu}, \citenamefont {Yu}, \citenamefont
  {Fang}, \citenamefont {Dai},\ and\ \citenamefont
  {Kawazoe}}]{weng_topological_2015}%
  \BibitemOpen
  \bibfield  {author} {\bibinfo {author} {\bibfnamefont {H.}~\bibnamefont
  {Weng}}, \bibinfo {author} {\bibfnamefont {Y.}~\bibnamefont {Liang}},
  \bibinfo {author} {\bibfnamefont {Q.}~\bibnamefont {Xu}}, \bibinfo {author}
  {\bibfnamefont {R.}~\bibnamefont {Yu}}, \bibinfo {author} {\bibfnamefont
  {Z.}~\bibnamefont {Fang}}, \bibinfo {author} {\bibfnamefont {X.}~\bibnamefont
  {Dai}},\ and\ \bibinfo {author} {\bibfnamefont {Y.}~\bibnamefont {Kawazoe}},\
  }\bibfield  {title} {\bibinfo {title} {Topological node-line semimetal in
  three-dimensional graphene networks},\ }\href
  {https://doi.org/10.1103/PhysRevB.92.045108} {\bibfield  {journal} {\bibinfo
  {journal} {Phys. Rev. B}\ }\textbf {\bibinfo {volume} {92}},\ \bibinfo
  {pages} {045108} (\bibinfo {year} {2015})}\BibitemShut {NoStop}%
\bibitem [{\citenamefont {Chen}\ \emph {et~al.}(2015)\citenamefont {Chen},
  \citenamefont {Xie}, \citenamefont {Yang}, \citenamefont {Pan}, \citenamefont
  {Zhang}, \citenamefont {Cohen},\ and\ \citenamefont {Zhang}}]{Chen2015}%
  \BibitemOpen
  \bibfield  {author} {\bibinfo {author} {\bibfnamefont {Y.}~\bibnamefont
  {Chen}}, \bibinfo {author} {\bibfnamefont {Y.}~\bibnamefont {Xie}}, \bibinfo
  {author} {\bibfnamefont {S.~A.}\ \bibnamefont {Yang}}, \bibinfo {author}
  {\bibfnamefont {H.}~\bibnamefont {Pan}}, \bibinfo {author} {\bibfnamefont
  {F.}~\bibnamefont {Zhang}}, \bibinfo {author} {\bibfnamefont {M.~L.}\
  \bibnamefont {Cohen}},\ and\ \bibinfo {author} {\bibfnamefont
  {S.}~\bibnamefont {Zhang}},\ }\bibfield  {title} {\bibinfo {title}
  {Nanostructured carbon allotropes with weyl-like loops and points},\ }\href
  {https://doi.org/10.1021/acs.nanolett.5b02978} {\bibfield  {journal}
  {\bibinfo  {journal} {Nano Lett.}\ }\textbf {\bibinfo {volume} {15}},\
  \bibinfo {pages} {6974} (\bibinfo {year} {2015})}\BibitemShut {NoStop}%
\bibitem [{\citenamefont {Yu}\ \emph {et~al.}(2015)\citenamefont {Yu},
  \citenamefont {Weng}, \citenamefont {Fang}, \citenamefont {Dai},\ and\
  \citenamefont {Hu}}]{Yu2015}%
  \BibitemOpen
  \bibfield  {author} {\bibinfo {author} {\bibfnamefont {R.}~\bibnamefont
  {Yu}}, \bibinfo {author} {\bibfnamefont {H.}~\bibnamefont {Weng}}, \bibinfo
  {author} {\bibfnamefont {Z.}~\bibnamefont {Fang}}, \bibinfo {author}
  {\bibfnamefont {X.}~\bibnamefont {Dai}},\ and\ \bibinfo {author}
  {\bibfnamefont {X.}~\bibnamefont {Hu}},\ }\bibfield  {title} {\bibinfo
  {title} {Topological node-line semimetal and dirac semimetal state in
  antiperovskite ${\mathrm{cu}}_{3}\mathrm{PdN}$},\ }\href
  {https://doi.org/10.1103/PhysRevLett.115.036807} {\bibfield  {journal}
  {\bibinfo  {journal} {Phys. Rev. Lett.}\ }\textbf {\bibinfo {volume} {115}},\
  \bibinfo {pages} {036807} (\bibinfo {year} {2015})}\BibitemShut {NoStop}%
\bibitem [{\citenamefont {Kim}\ \emph {et~al.}(2015)\citenamefont {Kim},
  \citenamefont {Wieder}, \citenamefont {Kane},\ and\ \citenamefont
  {Rappe}}]{Kim2015Dirac-Prl}%
  \BibitemOpen
  \bibfield  {author} {\bibinfo {author} {\bibfnamefont {Y.}~\bibnamefont
  {Kim}}, \bibinfo {author} {\bibfnamefont {B.~J.}\ \bibnamefont {Wieder}},
  \bibinfo {author} {\bibfnamefont {C.~L.}\ \bibnamefont {Kane}},\ and\
  \bibinfo {author} {\bibfnamefont {A.~M.}\ \bibnamefont {Rappe}},\ }\bibfield
  {title} {\bibinfo {title} {Dirac line nodes in inversion-symmetric
  crystals},\ }\href
  {https://journals.aps.org/prl/abstract/10.1103/PhysRevLett.115.036806}
  {\bibfield  {journal} {\bibinfo  {journal} {Physical review letters}\
  }\textbf {\bibinfo {volume} {115}},\ \bibinfo {pages} {036806} (\bibinfo
  {year} {2015})}\BibitemShut {NoStop}%
\bibitem [{\citenamefont {Mullen}\ \emph {et~al.}(2015)\citenamefont {Mullen},
  \citenamefont {Uchoa},\ and\ \citenamefont {Glatzhofer}}]{MullenPRL_2015}%
  \BibitemOpen
  \bibfield  {author} {\bibinfo {author} {\bibfnamefont {K.}~\bibnamefont
  {Mullen}}, \bibinfo {author} {\bibfnamefont {B.}~\bibnamefont {Uchoa}},\ and\
  \bibinfo {author} {\bibfnamefont {D.~T.}\ \bibnamefont {Glatzhofer}},\
  }\bibfield  {title} {\bibinfo {title} {Line of dirac nodes in hyperhoneycomb
  lattices},\ }\href {https://doi.org/10.1103/PhysRevLett.115.026403}
  {\bibfield  {journal} {\bibinfo  {journal} {Phys. Rev. Lett.}\ }\textbf
  {\bibinfo {volume} {115}},\ \bibinfo {pages} {026403} (\bibinfo {year}
  {2015})}\BibitemShut {NoStop}%
\bibitem [{\citenamefont {Liang}\ \emph {et~al.}(2016)\citenamefont {Liang},
  \citenamefont {Zhou}, \citenamefont {Yu}, \citenamefont {Wang},\ and\
  \citenamefont {Weng}}]{Liang2016Node-PRB}%
  \BibitemOpen
  \bibfield  {author} {\bibinfo {author} {\bibfnamefont {Q.-F.}\ \bibnamefont
  {Liang}}, \bibinfo {author} {\bibfnamefont {J.}~\bibnamefont {Zhou}},
  \bibinfo {author} {\bibfnamefont {R.}~\bibnamefont {Yu}}, \bibinfo {author}
  {\bibfnamefont {Z.}~\bibnamefont {Wang}},\ and\ \bibinfo {author}
  {\bibfnamefont {H.}~\bibnamefont {Weng}},\ }\bibfield  {title} {\bibinfo
  {title} {Node-surface and node-line fermions from nonsymmorphic lattice
  symmetries},\ }\href {https://link.aps.org/doi/10.1103/PhysRevB.93.085427}
  {\bibfield  {journal} {\bibinfo  {journal} {Phys. Rev. B}\ }\textbf {\bibinfo
  {volume} {93}},\ \bibinfo {pages} {085427} (\bibinfo {year}
  {2016})}\BibitemShut {NoStop}%
\bibitem [{\citenamefont {Zhong}\ \emph {et~al.}(2016)\citenamefont {Zhong},
  \citenamefont {Chen}, \citenamefont {Xie}, \citenamefont {Yang},
  \citenamefont {Cohen},\ and\ \citenamefont {Zhang}}]{Zhong_NS_2016}%
  \BibitemOpen
  \bibfield  {author} {\bibinfo {author} {\bibfnamefont {C.}~\bibnamefont
  {Zhong}}, \bibinfo {author} {\bibfnamefont {Y.}~\bibnamefont {Chen}},
  \bibinfo {author} {\bibfnamefont {Y.}~\bibnamefont {Xie}}, \bibinfo {author}
  {\bibfnamefont {S.~A.}\ \bibnamefont {Yang}}, \bibinfo {author}
  {\bibfnamefont {M.~L.}\ \bibnamefont {Cohen}},\ and\ \bibinfo {author}
  {\bibfnamefont {S.~B.}\ \bibnamefont {Zhang}},\ }\bibfield  {title} {\bibinfo
  {title} {Towards three-dimensional {Weyl-surface} semimetals in graphene
  networks},\ }\href {https://doi.org/10.1039/C6NR00882H} {\bibfield  {journal}
  {\bibinfo  {journal} {Nanoscale}\ }\textbf {\bibinfo {volume} {8}},\ \bibinfo
  {pages} {7232} (\bibinfo {year} {2016})}\BibitemShut {NoStop}%
\bibitem [{\citenamefont {Wu}\ \emph {et~al.}(2018)\citenamefont {Wu},
  \citenamefont {Liu}, \citenamefont {Li}, \citenamefont {Zhong}, \citenamefont
  {Yu}, \citenamefont {Sheng}, \citenamefont {Zhao},\ and\ \citenamefont
  {Yang}}]{wu_nodal_2018}%
  \BibitemOpen
  \bibfield  {author} {\bibinfo {author} {\bibfnamefont {W.}~\bibnamefont
  {Wu}}, \bibinfo {author} {\bibfnamefont {Y.}~\bibnamefont {Liu}}, \bibinfo
  {author} {\bibfnamefont {S.}~\bibnamefont {Li}}, \bibinfo {author}
  {\bibfnamefont {C.}~\bibnamefont {Zhong}}, \bibinfo {author} {\bibfnamefont
  {Z.-M.}\ \bibnamefont {Yu}}, \bibinfo {author} {\bibfnamefont {X.-L.}\
  \bibnamefont {Sheng}}, \bibinfo {author} {\bibfnamefont {Y.~X.}\ \bibnamefont
  {Zhao}},\ and\ \bibinfo {author} {\bibfnamefont {S.~A.}\ \bibnamefont
  {Yang}},\ }\bibfield  {title} {\bibinfo {title} {Nodal surface semimetals:
  {Theory} and material realization},\ }\href
  {https://doi.org/10.1103/PhysRevB.97.115125} {\bibfield  {journal} {\bibinfo
  {journal} {Phys. Rev. B}\ }\textbf {\bibinfo {volume} {97}},\ \bibinfo
  {pages} {115125} (\bibinfo {year} {2018})}\BibitemShut {NoStop}%
\bibitem [{\citenamefont {T{\"u}rker}\ and\ \citenamefont
  {Moroz}(2018)}]{turker_weyl_2018}%
  \BibitemOpen
  \bibfield  {author} {\bibinfo {author} {\bibfnamefont {O.}~\bibnamefont
  {T{\"u}rker}}\ and\ \bibinfo {author} {\bibfnamefont {S.}~\bibnamefont
  {Moroz}},\ }\bibfield  {title} {\bibinfo {title} {Weyl nodal surfaces},\
  }\href {https://doi.org/10.1103/PhysRevB.97.075120} {\bibfield  {journal}
  {\bibinfo  {journal} {Phys. Rev. B}\ }\textbf {\bibinfo {volume} {97}},\
  \bibinfo {pages} {075120} (\bibinfo {year} {2018})}\BibitemShut {NoStop}%
\bibitem [{\citenamefont {Zhang}\ \emph {et~al.}(2018)\citenamefont {Zhang},
  \citenamefont {Yu}, \citenamefont {Zhu}, \citenamefont {Wu}, \citenamefont
  {Wang}, \citenamefont {Sheng},\ and\ \citenamefont
  {Yang}}]{Zhang2018Nodal-PRB}%
  \BibitemOpen
  \bibfield  {author} {\bibinfo {author} {\bibfnamefont {X.}~\bibnamefont
  {Zhang}}, \bibinfo {author} {\bibfnamefont {Z.-M.}\ \bibnamefont {Yu}},
  \bibinfo {author} {\bibfnamefont {Z.}~\bibnamefont {Zhu}}, \bibinfo {author}
  {\bibfnamefont {W.}~\bibnamefont {Wu}}, \bibinfo {author} {\bibfnamefont
  {S.-S.}\ \bibnamefont {Wang}}, \bibinfo {author} {\bibfnamefont {X.-L.}\
  \bibnamefont {Sheng}},\ and\ \bibinfo {author} {\bibfnamefont {S.~A.}\
  \bibnamefont {Yang}},\ }\bibfield  {title} {\bibinfo {title} {Nodal loop and
  nodal surface states in the {Ti$_{3}$Al} family of materials},\ }\href
  {https://link.aps.org/doi/10.1103/PhysRevB.97.235150} {\bibfield  {journal}
  {\bibinfo  {journal} {Phys. Rev. B}\ }\textbf {\bibinfo {volume} {97}},\
  \bibinfo {pages} {235150} (\bibinfo {year} {2018})}\BibitemShut {NoStop}%
\bibitem [{\citenamefont {Xu}\ \emph {et~al.}(2011)\citenamefont {Xu},
  \citenamefont {Weng}, \citenamefont {Wang}, \citenamefont {Dai},\ and\
  \citenamefont {Fang}}]{Xu2011Chern-Prl}%
  \BibitemOpen
  \bibfield  {author} {\bibinfo {author} {\bibfnamefont {G.}~\bibnamefont
  {Xu}}, \bibinfo {author} {\bibfnamefont {H.}~\bibnamefont {Weng}}, \bibinfo
  {author} {\bibfnamefont {Z.}~\bibnamefont {Wang}}, \bibinfo {author}
  {\bibfnamefont {X.}~\bibnamefont {Dai}},\ and\ \bibinfo {author}
  {\bibfnamefont {Z.}~\bibnamefont {Fang}},\ }\bibfield  {title} {\bibinfo
  {title} {Chern semimetal and the quantized anomalous {Hall} effect in
  {HgCr$_2$Se$_4$}},\ }\href
  {https://link.aps.org/doi/10.1103/PhysRevLett.107.186806} {\bibfield
  {journal} {\bibinfo  {journal} {Phys. Rev. Lett.}\ }\textbf {\bibinfo
  {volume} {107}},\ \bibinfo {pages} {186806} (\bibinfo {year}
  {2011})}\BibitemShut {NoStop}%
\bibitem [{\citenamefont {Fang}\ \emph {et~al.}(2012)\citenamefont {Fang},
  \citenamefont {Gilbert}, \citenamefont {Dai},\ and\ \citenamefont
  {Bernevig}}]{fang_multi-weyl_2012}%
  \BibitemOpen
  \bibfield  {author} {\bibinfo {author} {\bibfnamefont {C.}~\bibnamefont
  {Fang}}, \bibinfo {author} {\bibfnamefont {M.~J.}\ \bibnamefont {Gilbert}},
  \bibinfo {author} {\bibfnamefont {X.}~\bibnamefont {Dai}},\ and\ \bibinfo
  {author} {\bibfnamefont {B.~A.}\ \bibnamefont {Bernevig}},\ }\bibfield
  {title} {\bibinfo {title} {Multi-{Weyl} {Topological} {Semimetals}
  {Stabilized} by {Point} {Group} {Symmetry}},\ }\href
  {https://doi.org/10.1103/PhysRevLett.108.266802} {\bibfield  {journal}
  {\bibinfo  {journal} {Phys. Rev. Lett.}\ }\textbf {\bibinfo {volume} {108}},\
  \bibinfo {pages} {266802} (\bibinfo {year} {2012})}\BibitemShut {NoStop}%
\bibitem [{\citenamefont {Liu}\ and\ \citenamefont
  {Zunger}(2017)}]{liu_predicted_2017}%
  \BibitemOpen
  \bibfield  {author} {\bibinfo {author} {\bibfnamefont {Q.}~\bibnamefont
  {Liu}}\ and\ \bibinfo {author} {\bibfnamefont {A.}~\bibnamefont {Zunger}},\
  }\bibfield  {title} {\bibinfo {title} {Predicted {Realization} of {Cubic}
  {Dirac} {Fermion} in {Quasi}-{One}-{Dimensional} {Transition}-{Metal}
  {Monochalcogenides}},\ }\href {https://doi.org/10.1103/PhysRevX.7.021019}
  {\bibfield  {journal} {\bibinfo  {journal} {Phys. Rev. X}\ }\textbf {\bibinfo
  {volume} {7}},\ \bibinfo {pages} {021019} (\bibinfo {year}
  {2017})}\BibitemShut {NoStop}%
\bibitem [{\citenamefont {He}\ \emph {et~al.}(2020{\natexlab{a}})\citenamefont
  {He}, \citenamefont {Qiu}, \citenamefont {Cai}, \citenamefont {Xiao},
  \citenamefont {Ke}, \citenamefont {Zhang},\ and\ \citenamefont
  {Liu}}]{he_observation_2020}%
  \BibitemOpen
  \bibfield  {author} {\bibinfo {author} {\bibfnamefont {H.}~\bibnamefont
  {He}}, \bibinfo {author} {\bibfnamefont {C.}~\bibnamefont {Qiu}}, \bibinfo
  {author} {\bibfnamefont {X.}~\bibnamefont {Cai}}, \bibinfo {author}
  {\bibfnamefont {M.}~\bibnamefont {Xiao}}, \bibinfo {author} {\bibfnamefont
  {M.}~\bibnamefont {Ke}}, \bibinfo {author} {\bibfnamefont {F.}~\bibnamefont
  {Zhang}},\ and\ \bibinfo {author} {\bibfnamefont {Z.}~\bibnamefont {Liu}},\
  }\bibfield  {title} {\bibinfo {title} {Observation of quadratic {Weyl} points
  and double-helicoid arcs},\ }\href
  {https://doi.org/10.1038/s41467-020-15825-5} {\bibfield  {journal} {\bibinfo
  {journal} {Nat. Commun.}\ }\textbf {\bibinfo {volume} {11}},\ \bibinfo
  {pages} {1820} (\bibinfo {year} {2020}{\natexlab{a}})}\BibitemShut {NoStop}%
\bibitem [{\citenamefont {Yang}\ and\ \citenamefont
  {Nagaosa}(2014)}]{yang_classification_2014}%
  \BibitemOpen
  \bibfield  {author} {\bibinfo {author} {\bibfnamefont {B.-J.}\ \bibnamefont
  {Yang}}\ and\ \bibinfo {author} {\bibfnamefont {N.}~\bibnamefont {Nagaosa}},\
  }\bibfield  {title} {\bibinfo {title} {Classification of stable
  three-dimensional {Dirac} semimetals with nontrivial topology},\ }\href
  {https://doi.org/10.1038/ncomms5898} {\bibfield  {journal} {\bibinfo
  {journal} {Nat. Commun.}\ }\textbf {\bibinfo {volume} {5}},\ \bibinfo {pages}
  {5898} (\bibinfo {year} {2014})}\BibitemShut {NoStop}%
\bibitem [{\citenamefont {Gao}\ \emph {et~al.}(2016)\citenamefont {Gao},
  \citenamefont {Hua}, \citenamefont {Zhang},\ and\ \citenamefont
  {Zhang}}]{Gao2016Classification-PRB}%
  \BibitemOpen
  \bibfield  {author} {\bibinfo {author} {\bibfnamefont {Z.}~\bibnamefont
  {Gao}}, \bibinfo {author} {\bibfnamefont {M.}~\bibnamefont {Hua}}, \bibinfo
  {author} {\bibfnamefont {H.}~\bibnamefont {Zhang}},\ and\ \bibinfo {author}
  {\bibfnamefont {X.}~\bibnamefont {Zhang}},\ }\bibfield  {title} {\bibinfo
  {title} {Classification of stable {Dirac} and {Weyl} semimetals with
  reflection and rotational symmetry},\ }\href
  {https://journals.aps.org/prb/abstract/10.1103/PhysRevB.93.205109} {\bibfield
   {journal} {\bibinfo  {journal} {Phys. Rev. B}\ }\textbf {\bibinfo {volume}
  {93}},\ \bibinfo {pages} {205109} (\bibinfo {year} {2016})}\BibitemShut
  {NoStop}%
\bibitem [{\citenamefont {Yu}\ \emph {et~al.}(2018)\citenamefont {Yu},
  \citenamefont {Zhou}, \citenamefont {Chuang}, \citenamefont {Yang},
  \citenamefont {Lin},\ and\ \citenamefont {Bansil}}]{Yu2018Nonsymmorphic-PRM}%
  \BibitemOpen
  \bibfield  {author} {\bibinfo {author} {\bibfnamefont {W.~C.}\ \bibnamefont
  {Yu}}, \bibinfo {author} {\bibfnamefont {X.}~\bibnamefont {Zhou}}, \bibinfo
  {author} {\bibfnamefont {F.-C.}\ \bibnamefont {Chuang}}, \bibinfo {author}
  {\bibfnamefont {S.~A.}\ \bibnamefont {Yang}}, \bibinfo {author}
  {\bibfnamefont {H.}~\bibnamefont {Lin}},\ and\ \bibinfo {author}
  {\bibfnamefont {A.}~\bibnamefont {Bansil}},\ }\bibfield  {title} {\bibinfo
  {title} {Nonsymmorphic cubic {Dirac} point and crossed nodal rings across the
  ferroelectric phase transition in {LiOsO$_3$}},\ }\href
  {https://journals.aps.org/prmaterials/abstract/10.1103/PhysRevMaterials.2.051201}
  {\bibfield  {journal} {\bibinfo  {journal} {Phys. Rev. Mater.}\ }\textbf
  {\bibinfo {volume} {2}},\ \bibinfo {pages} {051201} (\bibinfo {year}
  {2018})}\BibitemShut {NoStop}%
\bibitem [{\citenamefont {Wu}\ \emph {et~al.}(2020)\citenamefont {Wu},
  \citenamefont {Yu}, \citenamefont {Zhou}, \citenamefont {Zhao},\ and\
  \citenamefont {Yang}}]{wu_higher-order_2020}%
  \BibitemOpen
  \bibfield  {author} {\bibinfo {author} {\bibfnamefont {W.}~\bibnamefont
  {Wu}}, \bibinfo {author} {\bibfnamefont {Z.-M.}\ \bibnamefont {Yu}}, \bibinfo
  {author} {\bibfnamefont {X.}~\bibnamefont {Zhou}}, \bibinfo {author}
  {\bibfnamefont {Y.~X.}\ \bibnamefont {Zhao}},\ and\ \bibinfo {author}
  {\bibfnamefont {S.~A.}\ \bibnamefont {Yang}},\ }\bibfield  {title} {\bibinfo
  {title} {Higher-order {Dirac} fermions in three dimensions},\ }\href
  {https://doi.org/10.1103/PhysRevB.101.205134} {\bibfield  {journal} {\bibinfo
   {journal} {Phys. Rev. B}\ }\textbf {\bibinfo {volume} {101}},\ \bibinfo
  {pages} {205134} (\bibinfo {year} {2020})}\BibitemShut {NoStop}%
\bibitem [{\citenamefont {Yu}\ \emph {et~al.}(2019{\natexlab{a}})\citenamefont
  {Yu}, \citenamefont {Wu}, \citenamefont {Sheng}, \citenamefont {Zhao},\ and\
  \citenamefont {Yang}}]{yu_quadratic_2019}%
  \BibitemOpen
  \bibfield  {author} {\bibinfo {author} {\bibfnamefont {Z.-M.}\ \bibnamefont
  {Yu}}, \bibinfo {author} {\bibfnamefont {W.}~\bibnamefont {Wu}}, \bibinfo
  {author} {\bibfnamefont {X.-L.}\ \bibnamefont {Sheng}}, \bibinfo {author}
  {\bibfnamefont {Y.~X.}\ \bibnamefont {Zhao}},\ and\ \bibinfo {author}
  {\bibfnamefont {S.~A.}\ \bibnamefont {Yang}},\ }\bibfield  {title} {\bibinfo
  {title} {Quadratic and cubic nodal lines stabilized by crystalline
  symmetry},\ }\href {https://doi.org/10.1103/PhysRevB.99.121106} {\bibfield
  {journal} {\bibinfo  {journal} {Phys. Rev. B}\ }\textbf {\bibinfo {volume}
  {99}},\ \bibinfo {pages} {121106(R)} (\bibinfo {year}
  {2019}{\natexlab{a}})}\BibitemShut {NoStop}%
\bibitem [{\citenamefont {Wang}\ \emph {et~al.}(2020)\citenamefont {Wang},
  \citenamefont {Li},\ and\ \citenamefont {Zhang}}]{wang_possible_2020}%
  \BibitemOpen
  \bibfield  {author} {\bibinfo {author} {\bibfnamefont {J.-R.}\ \bibnamefont
  {Wang}}, \bibinfo {author} {\bibfnamefont {W.}~\bibnamefont {Li}},\ and\
  \bibinfo {author} {\bibfnamefont {C.-J.}\ \bibnamefont {Zhang}},\ }\bibfield
  {title} {\bibinfo {title} {Possible instabilities in quadratic and cubic
  nodal-line fermion systems with correlated interactions},\ }\href
  {https://doi.org/10.1103/PhysRevB.102.085132} {\bibfield  {journal} {\bibinfo
   {journal} {Phys. Rev. B}\ }\textbf {\bibinfo {volume} {102}},\ \bibinfo
  {pages} {085132} (\bibinfo {year} {2020})}\BibitemShut {NoStop}%
\bibitem [{\citenamefont {Morali}\ \emph {et~al.}(2019)\citenamefont {Morali},
  \citenamefont {Batabyal}, \citenamefont {Nag}, \citenamefont {Liu},
  \citenamefont {Xu}, \citenamefont {Sun}, \citenamefont {Yan}, \citenamefont
  {Felser}, \citenamefont {Avraham},\ and\ \citenamefont
  {Beidenkopf}}]{morali2019fermi}%
  \BibitemOpen
  \bibfield  {author} {\bibinfo {author} {\bibfnamefont {N.}~\bibnamefont
  {Morali}}, \bibinfo {author} {\bibfnamefont {R.}~\bibnamefont {Batabyal}},
  \bibinfo {author} {\bibfnamefont {P.~K.}\ \bibnamefont {Nag}}, \bibinfo
  {author} {\bibfnamefont {E.}~\bibnamefont {Liu}}, \bibinfo {author}
  {\bibfnamefont {Q.}~\bibnamefont {Xu}}, \bibinfo {author} {\bibfnamefont
  {Y.}~\bibnamefont {Sun}}, \bibinfo {author} {\bibfnamefont {B.}~\bibnamefont
  {Yan}}, \bibinfo {author} {\bibfnamefont {C.}~\bibnamefont {Felser}},
  \bibinfo {author} {\bibfnamefont {N.}~\bibnamefont {Avraham}},\ and\ \bibinfo
  {author} {\bibfnamefont {H.}~\bibnamefont {Beidenkopf}},\ }\bibfield  {title}
  {\bibinfo {title} {Fermi-arc diversity on surface terminations of the
  magnetic {Weyl} semimetal {Co$_3$Sn$_2$S$_2$}},\ }\href
  {https://science.sciencemag.org/content/365/6459/1286.full} {\bibfield
  {journal} {\bibinfo  {journal} {Science}\ }\textbf {\bibinfo {volume}
  {365}},\ \bibinfo {pages} {1286} (\bibinfo {year} {2019})}\BibitemShut
  {NoStop}%
\bibitem [{\citenamefont {You}\ \emph {et~al.}(2019)\citenamefont {You},
  \citenamefont {Chen}, \citenamefont {Zhang}, \citenamefont {Sheng},
  \citenamefont {Yang},\ and\ \citenamefont {Su}}]{You2019}%
  \BibitemOpen
  \bibfield  {author} {\bibinfo {author} {\bibfnamefont {J.-Y.}\ \bibnamefont
  {You}}, \bibinfo {author} {\bibfnamefont {C.}~\bibnamefont {Chen}}, \bibinfo
  {author} {\bibfnamefont {Z.}~\bibnamefont {Zhang}}, \bibinfo {author}
  {\bibfnamefont {X.-L.}\ \bibnamefont {Sheng}}, \bibinfo {author}
  {\bibfnamefont {S.~A.}\ \bibnamefont {Yang}},\ and\ \bibinfo {author}
  {\bibfnamefont {G.}~\bibnamefont {Su}},\ }\bibfield  {title} {\bibinfo
  {title} {Two-dimensional weyl half-semimetal and tunable quantum anomalous
  hall effect},\ }\href {https://doi.org/10.1103/PhysRevB.100.064408}
  {\bibfield  {journal} {\bibinfo  {journal} {Phys. Rev. B}\ }\textbf {\bibinfo
  {volume} {100}},\ \bibinfo {pages} {064408} (\bibinfo {year}
  {2019})}\BibitemShut {NoStop}%
\bibitem [{\citenamefont {Liu}\ \emph {et~al.}(2019)\citenamefont {Liu},
  \citenamefont {Liang}, \citenamefont {Liu}, \citenamefont {Xu}, \citenamefont
  {Li}, \citenamefont {Chen}, \citenamefont {Pei}, \citenamefont {Shi},
  \citenamefont {Mo}, \citenamefont {Dudin} \emph {et~al.}}]{liu2019magnetic}%
  \BibitemOpen
  \bibfield  {author} {\bibinfo {author} {\bibfnamefont {D.}~\bibnamefont
  {Liu}}, \bibinfo {author} {\bibfnamefont {A.}~\bibnamefont {Liang}}, \bibinfo
  {author} {\bibfnamefont {E.}~\bibnamefont {Liu}}, \bibinfo {author}
  {\bibfnamefont {Q.}~\bibnamefont {Xu}}, \bibinfo {author} {\bibfnamefont
  {Y.}~\bibnamefont {Li}}, \bibinfo {author} {\bibfnamefont {C.}~\bibnamefont
  {Chen}}, \bibinfo {author} {\bibfnamefont {D.}~\bibnamefont {Pei}}, \bibinfo
  {author} {\bibfnamefont {W.}~\bibnamefont {Shi}}, \bibinfo {author}
  {\bibfnamefont {S.}~\bibnamefont {Mo}}, \bibinfo {author} {\bibfnamefont
  {P.}~\bibnamefont {Dudin}}, \emph {et~al.},\ }\bibfield  {title} {\bibinfo
  {title} {Magnetic weyl semimetal phase in a kagom{\'e} crystal},\ }\href
  {https://science.sciencemag.org/content/365/6459/1282.full} {\bibfield
  {journal} {\bibinfo  {journal} {Science}\ }\textbf {\bibinfo {volume}
  {365}},\ \bibinfo {pages} {1282} (\bibinfo {year} {2019})}\BibitemShut
  {NoStop}%
\bibitem [{\citenamefont {Belopolski}\ \emph {et~al.}(2019)\citenamefont
  {Belopolski}, \citenamefont {Manna}, \citenamefont {Sanchez}, \citenamefont
  {Chang}, \citenamefont {Ernst}, \citenamefont {Yin}, \citenamefont {Zhang},
  \citenamefont {Cochran}, \citenamefont {Shumiya}, \citenamefont {Zheng} \emph
  {et~al.}}]{belopolski2019discovery}%
  \BibitemOpen
  \bibfield  {author} {\bibinfo {author} {\bibfnamefont {I.}~\bibnamefont
  {Belopolski}}, \bibinfo {author} {\bibfnamefont {K.}~\bibnamefont {Manna}},
  \bibinfo {author} {\bibfnamefont {D.~S.}\ \bibnamefont {Sanchez}}, \bibinfo
  {author} {\bibfnamefont {G.}~\bibnamefont {Chang}}, \bibinfo {author}
  {\bibfnamefont {B.}~\bibnamefont {Ernst}}, \bibinfo {author} {\bibfnamefont
  {J.}~\bibnamefont {Yin}}, \bibinfo {author} {\bibfnamefont {S.~S.}\
  \bibnamefont {Zhang}}, \bibinfo {author} {\bibfnamefont {T.}~\bibnamefont
  {Cochran}}, \bibinfo {author} {\bibfnamefont {N.}~\bibnamefont {Shumiya}},
  \bibinfo {author} {\bibfnamefont {H.}~\bibnamefont {Zheng}}, \emph {et~al.},\
  }\bibfield  {title} {\bibinfo {title} {Discovery of topological {Weyl}
  fermion lines and drumhead surface states in a room temperature magnet},\
  }\href {https://science.sciencemag.org/content/365/6459/1278.full} {\bibfield
   {journal} {\bibinfo  {journal} {Science}\ }\textbf {\bibinfo {volume}
  {365}},\ \bibinfo {pages} {1278} (\bibinfo {year} {2019})}\BibitemShut
  {NoStop}%
\bibitem [{\citenamefont {Xu}\ \emph {et~al.}(2020)\citenamefont {Xu},
  \citenamefont {Elcoro}, \citenamefont {Song}, \citenamefont {Wieder},
  \citenamefont {Vergniory}, \citenamefont {Regnault}, \citenamefont {Chen},
  \citenamefont {Felser},\ and\ \citenamefont {Bernevig}}]{xu2020high}%
  \BibitemOpen
  \bibfield  {author} {\bibinfo {author} {\bibfnamefont {Y.}~\bibnamefont
  {Xu}}, \bibinfo {author} {\bibfnamefont {L.}~\bibnamefont {Elcoro}}, \bibinfo
  {author} {\bibfnamefont {Z.-D.}\ \bibnamefont {Song}}, \bibinfo {author}
  {\bibfnamefont {B.~J.}\ \bibnamefont {Wieder}}, \bibinfo {author}
  {\bibfnamefont {M.}~\bibnamefont {Vergniory}}, \bibinfo {author}
  {\bibfnamefont {N.}~\bibnamefont {Regnault}}, \bibinfo {author}
  {\bibfnamefont {Y.}~\bibnamefont {Chen}}, \bibinfo {author} {\bibfnamefont
  {C.}~\bibnamefont {Felser}},\ and\ \bibinfo {author} {\bibfnamefont {B.~A.}\
  \bibnamefont {Bernevig}},\ }\bibfield  {title} {\bibinfo {title}
  {High-throughput calculations of magnetic topological materials},\ }\href
  {https://www.nature.com/articles/s41586-020-2837-0} {\bibfield  {journal}
  {\bibinfo  {journal} {Nature}\ }\textbf {\bibinfo {volume} {586}},\ \bibinfo
  {pages} {702} (\bibinfo {year} {2020})}\BibitemShut {NoStop}%
\bibitem [{\citenamefont {Jin}\ \emph {et~al.}(2020)\citenamefont {Jin},
  \citenamefont {Zhang}, \citenamefont {Liu}, \citenamefont {Dai},
  \citenamefont {Wang},\ and\ \citenamefont {Liu}}]{Jin_double_PRB2020}%
  \BibitemOpen
  \bibfield  {author} {\bibinfo {author} {\bibfnamefont {L.}~\bibnamefont
  {Jin}}, \bibinfo {author} {\bibfnamefont {X.}~\bibnamefont {Zhang}}, \bibinfo
  {author} {\bibfnamefont {Y.}~\bibnamefont {Liu}}, \bibinfo {author}
  {\bibfnamefont {X.}~\bibnamefont {Dai}}, \bibinfo {author} {\bibfnamefont
  {L.}~\bibnamefont {Wang}},\ and\ \bibinfo {author} {\bibfnamefont
  {G.}~\bibnamefont {Liu}},\ }\bibfield  {title} {\bibinfo {title} {Fully
  spin-polarized double-weyl fermions with {type-III} dispersion in the
  quasi-one-dimensional materials {X$_{2}$RhF$_{6}$ (X$=$K, Rb, Cs)}},\ }\href
  {https://doi.org/10.1103/PhysRevB.102.195104} {\bibfield  {journal} {\bibinfo
   {journal} {Phys. Rev. B}\ }\textbf {\bibinfo {volume} {102}},\ \bibinfo
  {pages} {195104} (\bibinfo {year} {2020})}\BibitemShut {NoStop}%
\bibitem [{\citenamefont {Puphal}\ \emph {et~al.}(2020)\citenamefont {Puphal},
  \citenamefont {Pomjakushin}, \citenamefont {Kanazawa}, \citenamefont
  {Ukleev}, \citenamefont {Gawryluk}, \citenamefont {Ma}, \citenamefont
  {Naamneh}, \citenamefont {Plumb}, \citenamefont {Keller}, \citenamefont
  {Cubitt}, \citenamefont {Pomjakushina},\ and\ \citenamefont
  {White}}]{Mag_Puphal_PRL2020}%
  \BibitemOpen
  \bibfield  {author} {\bibinfo {author} {\bibfnamefont {P.}~\bibnamefont
  {Puphal}}, \bibinfo {author} {\bibfnamefont {V.}~\bibnamefont {Pomjakushin}},
  \bibinfo {author} {\bibfnamefont {N.}~\bibnamefont {Kanazawa}}, \bibinfo
  {author} {\bibfnamefont {V.}~\bibnamefont {Ukleev}}, \bibinfo {author}
  {\bibfnamefont {D.~J.}\ \bibnamefont {Gawryluk}}, \bibinfo {author}
  {\bibfnamefont {J.}~\bibnamefont {Ma}}, \bibinfo {author} {\bibfnamefont
  {M.}~\bibnamefont {Naamneh}}, \bibinfo {author} {\bibfnamefont {N.~C.}\
  \bibnamefont {Plumb}}, \bibinfo {author} {\bibfnamefont {L.}~\bibnamefont
  {Keller}}, \bibinfo {author} {\bibfnamefont {R.}~\bibnamefont {Cubitt}},
  \bibinfo {author} {\bibfnamefont {E.}~\bibnamefont {Pomjakushina}},\ and\
  \bibinfo {author} {\bibfnamefont {J.~S.}\ \bibnamefont {White}},\ }\bibfield
  {title} {\bibinfo {title} {Topological magnetic phase in the candidate weyl
  semimetal cealge},\ }\href {https://doi.org/10.1103/PhysRevLett.124.017202}
  {\bibfield  {journal} {\bibinfo  {journal} {Phys. Rev. Lett.}\ }\textbf
  {\bibinfo {volume} {124}},\ \bibinfo {pages} {017202} (\bibinfo {year}
  {2020})}\BibitemShut {NoStop}%
\bibitem [{\citenamefont {Zou}\ \emph {et~al.}(2019)\citenamefont {Zou},
  \citenamefont {He},\ and\ \citenamefont {Xu}}]{Mag_XGang_npj2020}%
  \BibitemOpen
  \bibfield  {author} {\bibinfo {author} {\bibfnamefont {J.}~\bibnamefont
  {Zou}}, \bibinfo {author} {\bibfnamefont {Z.}~\bibnamefont {He}},\ and\
  \bibinfo {author} {\bibfnamefont {G.}~\bibnamefont {Xu}},\ }\bibfield
  {title} {\bibinfo {title} {The study of magnetic topological semimetals by
  first principles calculations},\ }\href
  {https://doi.org/10.1038/s41524-019-0237-5} {\bibfield  {journal} {\bibinfo
  {journal} {npj Computational Materials}\ }\textbf {\bibinfo {volume} {5}},\
  \bibinfo {pages} {96} (\bibinfo {year} {2019})}\BibitemShut {NoStop}%
\bibitem [{\citenamefont {Wang}(2017)}]{wang_antiferromagnetic_2017}%
  \BibitemOpen
  \bibfield  {author} {\bibinfo {author} {\bibfnamefont {J.}~\bibnamefont
  {Wang}},\ }\bibfield  {title} {\bibinfo {title} {Antiferromagnetic
  topological nodal line semimetals},\ }\href
  {https://doi.org/10.1103/PhysRevB.96.081107} {\bibfield  {journal} {\bibinfo
  {journal} {Phys. Rev. B}\ }\textbf {\bibinfo {volume} {96}},\ \bibinfo
  {pages} {081107} (\bibinfo {year} {2017})}\BibitemShut {NoStop}%
\bibitem [{\citenamefont {Wang}\ \emph {et~al.}(2018)\citenamefont {Wang},
  \citenamefont {Gao}, \citenamefont {Lu}, \citenamefont {Xie}, \citenamefont
  {Ge}, \citenamefont {Zhao}, \citenamefont {Zhang},\ and\ \citenamefont
  {Liu}}]{Wang2018}%
  \BibitemOpen
  \bibfield  {author} {\bibinfo {author} {\bibfnamefont {B.}~\bibnamefont
  {Wang}}, \bibinfo {author} {\bibfnamefont {H.}~\bibnamefont {Gao}}, \bibinfo
  {author} {\bibfnamefont {Q.}~\bibnamefont {Lu}}, \bibinfo {author}
  {\bibfnamefont {W.}~\bibnamefont {Xie}}, \bibinfo {author} {\bibfnamefont
  {Y.}~\bibnamefont {Ge}}, \bibinfo {author} {\bibfnamefont {Y.-H.}\
  \bibnamefont {Zhao}}, \bibinfo {author} {\bibfnamefont {K.}~\bibnamefont
  {Zhang}},\ and\ \bibinfo {author} {\bibfnamefont {Y.}~\bibnamefont {Liu}},\
  }\bibfield  {title} {\bibinfo {title} {Type-i and type-ii nodal lines
  coexistence in the antiferromagnetic monolayer ${\mathrm{cras}}_{2}$},\
  }\href {https://doi.org/10.1103/PhysRevB.98.115164} {\bibfield  {journal}
  {\bibinfo  {journal} {Phys. Rev. B}\ }\textbf {\bibinfo {volume} {98}},\
  \bibinfo {pages} {115164} (\bibinfo {year} {2018})}\BibitemShut {NoStop}%
\bibitem [{\citenamefont {Wang}\ \emph {et~al.}(2019)\citenamefont {Wang},
  \citenamefont {Yu}, \citenamefont {Liu}, \citenamefont {Jiao}, \citenamefont
  {Guan}, \citenamefont {Sheng},\ and\ \citenamefont {Yang}}]{Wang2019}%
  \BibitemOpen
  \bibfield  {author} {\bibinfo {author} {\bibfnamefont {S.-S.}\ \bibnamefont
  {Wang}}, \bibinfo {author} {\bibfnamefont {Z.-M.}\ \bibnamefont {Yu}},
  \bibinfo {author} {\bibfnamefont {Y.}~\bibnamefont {Liu}}, \bibinfo {author}
  {\bibfnamefont {Y.}~\bibnamefont {Jiao}}, \bibinfo {author} {\bibfnamefont
  {S.}~\bibnamefont {Guan}}, \bibinfo {author} {\bibfnamefont {X.-L.}\
  \bibnamefont {Sheng}},\ and\ \bibinfo {author} {\bibfnamefont {S.~A.}\
  \bibnamefont {Yang}},\ }\bibfield  {title} {\bibinfo {title} {Two-dimensional
  nodal-loop half-metal in monolayer mnn},\ }\href
  {https://doi.org/10.1103/PhysRevMaterials.3.084201} {\bibfield  {journal}
  {\bibinfo  {journal} {Phys. Rev. Materials}\ }\textbf {\bibinfo {volume}
  {3}},\ \bibinfo {pages} {084201} (\bibinfo {year} {2019})}\BibitemShut
  {NoStop}%
\bibitem [{\citenamefont {Chen}\ \emph {et~al.}(2019)\citenamefont {Chen},
  \citenamefont {Yu}, \citenamefont {Li}, \citenamefont {Chen}, \citenamefont
  {Sheng},\ and\ \citenamefont {Yang}}]{Chen2019}%
  \BibitemOpen
  \bibfield  {author} {\bibinfo {author} {\bibfnamefont {C.}~\bibnamefont
  {Chen}}, \bibinfo {author} {\bibfnamefont {Z.-M.}\ \bibnamefont {Yu}},
  \bibinfo {author} {\bibfnamefont {S.}~\bibnamefont {Li}}, \bibinfo {author}
  {\bibfnamefont {Z.}~\bibnamefont {Chen}}, \bibinfo {author} {\bibfnamefont
  {X.-L.}\ \bibnamefont {Sheng}},\ and\ \bibinfo {author} {\bibfnamefont
  {S.~A.}\ \bibnamefont {Yang}},\ }\bibfield  {title} {\bibinfo {title}
  {Weyl-loop half-metal in ${\mathrm{li}}_{3}{({\mathrm{FeO}}_{3})}_{2}$},\
  }\href {https://doi.org/10.1103/PhysRevB.99.075131} {\bibfield  {journal}
  {\bibinfo  {journal} {Phys. Rev. B}\ }\textbf {\bibinfo {volume} {99}},\
  \bibinfo {pages} {075131} (\bibinfo {year} {2019})}\BibitemShut {NoStop}%
\bibitem [{\citenamefont {Feng}\ \emph {et~al.}(2019)\citenamefont {Feng},
  \citenamefont {Zhang}, \citenamefont {Feng}, \citenamefont {Fu},
  \citenamefont {Wu}, \citenamefont {Miyamoto}, \citenamefont {He},
  \citenamefont {Chen}, \citenamefont {Wu}, \citenamefont {Shimada},
  \citenamefont {Okuda},\ and\ \citenamefont {Yao}}]{Feng2019}%
  \BibitemOpen
  \bibfield  {author} {\bibinfo {author} {\bibfnamefont {B.}~\bibnamefont
  {Feng}}, \bibinfo {author} {\bibfnamefont {R.-W.}\ \bibnamefont {Zhang}},
  \bibinfo {author} {\bibfnamefont {Y.}~\bibnamefont {Feng}}, \bibinfo {author}
  {\bibfnamefont {B.}~\bibnamefont {Fu}}, \bibinfo {author} {\bibfnamefont
  {S.}~\bibnamefont {Wu}}, \bibinfo {author} {\bibfnamefont {K.}~\bibnamefont
  {Miyamoto}}, \bibinfo {author} {\bibfnamefont {S.}~\bibnamefont {He}},
  \bibinfo {author} {\bibfnamefont {L.}~\bibnamefont {Chen}}, \bibinfo {author}
  {\bibfnamefont {K.}~\bibnamefont {Wu}}, \bibinfo {author} {\bibfnamefont
  {K.}~\bibnamefont {Shimada}}, \bibinfo {author} {\bibfnamefont
  {T.}~\bibnamefont {Okuda}},\ and\ \bibinfo {author} {\bibfnamefont
  {Y.}~\bibnamefont {Yao}},\ }\bibfield  {title} {\bibinfo {title} {Discovery
  of weyl nodal lines in a single-layer ferromagnet},\ }\href
  {https://doi.org/10.1103/PhysRevLett.123.116401} {\bibfield  {journal}
  {\bibinfo  {journal} {Phys. Rev. Lett.}\ }\textbf {\bibinfo {volume} {123}},\
  \bibinfo {pages} {116401} (\bibinfo {year} {2019})}\BibitemShut {NoStop}%
\bibitem [{\citenamefont {Nie}\ \emph {et~al.}(2020)\citenamefont {Nie},
  \citenamefont {Sun}, \citenamefont {Prinz}, \citenamefont {Wang},
  \citenamefont {Weng}, \citenamefont {Fang},\ and\ \citenamefont
  {Dai}}]{nie_magnetic_2020}%
  \BibitemOpen
  \bibfield  {author} {\bibinfo {author} {\bibfnamefont {S.}~\bibnamefont
  {Nie}}, \bibinfo {author} {\bibfnamefont {Y.}~\bibnamefont {Sun}}, \bibinfo
  {author} {\bibfnamefont {F.~B.}\ \bibnamefont {Prinz}}, \bibinfo {author}
  {\bibfnamefont {Z.}~\bibnamefont {Wang}}, \bibinfo {author} {\bibfnamefont
  {H.}~\bibnamefont {Weng}}, \bibinfo {author} {\bibfnamefont {Z.}~\bibnamefont
  {Fang}},\ and\ \bibinfo {author} {\bibfnamefont {X.}~\bibnamefont {Dai}},\
  }\bibfield  {title} {\bibinfo {title} {Magnetic {Semimetals} and {Quantized}
  {Anomalous} {Hall} {Effect} in {EuB}$_{\textrm{6}}$},\ }\href
  {https://doi.org/10.1103/PhysRevLett.124.076403} {\bibfield  {journal}
  {\bibinfo  {journal} {Phys. Rev. Lett.}\ }\textbf {\bibinfo {volume} {124}},\
  \bibinfo {pages} {076403} (\bibinfo {year} {2020})}\BibitemShut {NoStop}%
\bibitem [{\citenamefont {He}\ \emph {et~al.}(2020{\natexlab{b}})\citenamefont
  {He}, \citenamefont {Zhang}, \citenamefont {Liu}, \citenamefont {Dai},
  \citenamefont {Liu}, \citenamefont {Yu},\ and\ \citenamefont
  {Yao}}]{Hezhang_NL_PRB2020}%
  \BibitemOpen
  \bibfield  {author} {\bibinfo {author} {\bibfnamefont {T.}~\bibnamefont
  {He}}, \bibinfo {author} {\bibfnamefont {X.}~\bibnamefont {Zhang}}, \bibinfo
  {author} {\bibfnamefont {Y.}~\bibnamefont {Liu}}, \bibinfo {author}
  {\bibfnamefont {X.}~\bibnamefont {Dai}}, \bibinfo {author} {\bibfnamefont
  {G.}~\bibnamefont {Liu}}, \bibinfo {author} {\bibfnamefont {Z.-M.}\
  \bibnamefont {Yu}},\ and\ \bibinfo {author} {\bibfnamefont {Y.}~\bibnamefont
  {Yao}},\ }\bibfield  {title} {\bibinfo {title} {Ferromagnetic hybrid nodal
  loop and switchable {type-I} and {type-II} weyl fermions in two dimensions},\
  }\href {https://doi.org/10.1103/PhysRevB.102.075133} {\bibfield  {journal}
  {\bibinfo  {journal} {Phys. Rev. B}\ }\textbf {\bibinfo {volume} {102}},\
  \bibinfo {pages} {075133} (\bibinfo {year} {2020}{\natexlab{b}})}\BibitemShut
  {NoStop}%
\bibitem [{\citenamefont {Song}\ and\ \citenamefont
  {Lee}(2020)}]{song_symmetry-protected_2020}%
  \BibitemOpen
  \bibfield  {author} {\bibinfo {author} {\bibfnamefont {Y.-J.}\ \bibnamefont
  {Song}}\ and\ \bibinfo {author} {\bibfnamefont {K.-W.}\ \bibnamefont {Lee}},\
  }\bibfield  {title} {\bibinfo {title} {Symmetry-protected spinful magnetic
  {Weyl} nodal loops and multi-{Weyl} nodes in $5d^n$ cubic double perovskites
  $(n=1,2)$},\ }\href {https://doi.org/10.1103/PhysRevB.102.035155} {\bibfield
  {journal} {\bibinfo  {journal} {Phys. Rev. B}\ }\textbf {\bibinfo {volume}
  {102}},\ \bibinfo {pages} {035155} (\bibinfo {year} {2020})}\BibitemShut
  {NoStop}%
\bibitem [{\citenamefont {Zhang}\ \emph {et~al.}(2020)\citenamefont {Zhang},
  \citenamefont {Zhang}, \citenamefont {Liu},\ and\ \citenamefont
  {Yao}}]{ZhangRW_NL_2020}%
  \BibitemOpen
  \bibfield  {author} {\bibinfo {author} {\bibfnamefont {R.-W.}\ \bibnamefont
  {Zhang}}, \bibinfo {author} {\bibfnamefont {Z.}~\bibnamefont {Zhang}},
  \bibinfo {author} {\bibfnamefont {C.-C.}\ \bibnamefont {Liu}},\ and\ \bibinfo
  {author} {\bibfnamefont {Y.}~\bibnamefont {Yao}},\ }\bibfield  {title}
  {\bibinfo {title} {Nodal line spin-gapless semimetals and high-quality
  candidate materials},\ }\href
  {https://doi.org/10.1103/PhysRevLett.124.016402} {\bibfield  {journal}
  {\bibinfo  {journal} {Phys. Rev. Lett.}\ }\textbf {\bibinfo {volume} {124}},\
  \bibinfo {pages} {016402} (\bibinfo {year} {2020})}\BibitemShut {NoStop}%
\bibitem [{\citenamefont {Wieder}\ and\ \citenamefont
  {Kane}(2016)}]{wieder_spin-orbit_2016}%
  \BibitemOpen
  \bibfield  {author} {\bibinfo {author} {\bibfnamefont {B.~J.}\ \bibnamefont
  {Wieder}}\ and\ \bibinfo {author} {\bibfnamefont {C.~L.}\ \bibnamefont
  {Kane}},\ }\bibfield  {title} {\bibinfo {title} {Spin-orbit semimetals in the
  layer groups},\ }\href {https://doi.org/10.1103/PhysRevB.94.155108}
  {\bibfield  {journal} {\bibinfo  {journal} {Phys. Rev. B}\ }\textbf {\bibinfo
  {volume} {94}},\ \bibinfo {pages} {155108} (\bibinfo {year}
  {2016})}\BibitemShut {NoStop}%
\bibitem [{\citenamefont {Yu}\ \emph {et~al.}(2019{\natexlab{b}})\citenamefont
  {Yu}, \citenamefont {Wu}, \citenamefont {Zhao},\ and\ \citenamefont
  {Yang}}]{Yu2019Circumventing-PRB}%
  \BibitemOpen
  \bibfield  {author} {\bibinfo {author} {\bibfnamefont {Z.-M.}\ \bibnamefont
  {Yu}}, \bibinfo {author} {\bibfnamefont {W.}~\bibnamefont {Wu}}, \bibinfo
  {author} {\bibfnamefont {Y.~X.}\ \bibnamefont {Zhao}},\ and\ \bibinfo
  {author} {\bibfnamefont {S.~A.}\ \bibnamefont {Yang}},\ }\bibfield  {title}
  {\bibinfo {title} {Circumventing the no-go theorem: A single {Weyl} point
  without surface fermi arcs},\ }\href
  {https://link.aps.org/doi/10.1103/PhysRevB.100.041118} {\bibfield  {journal}
  {\bibinfo  {journal} {Phys. Rev. B}\ }\textbf {\bibinfo {volume} {100}},\
  \bibinfo {pages} {041118} (\bibinfo {year} {2019}{\natexlab{b}})}\BibitemShut
  {NoStop}%
\bibitem [{\citenamefont {Bradley}\ and\ \citenamefont
  {Cracknell}()}]{bradley_mathematical_nodate}%
  \BibitemOpen
  \bibfield  {author} {\bibinfo {author} {\bibfnamefont {C.}~\bibnamefont
  {Bradley}}\ and\ \bibinfo {author} {\bibfnamefont {A.}~\bibnamefont
  {Cracknell}},\ }\bibfield  {title} {\bibinfo {title} {The {Mathematical}
  {Theory} of {Symmetry} in {Solids}: {Representation} {Theory} for {Point}
  {Groups} and {Space} {Groups}},\ }\href@noop {} {\bibinfo  {journal} {The
  Mathematical Theory of Symmetry in Solids: Representation Theory for Point
  Groups and Space Groups (Oxford University Press, New York, 1972)}\
  }\BibitemShut {NoStop}%
\bibitem [{\citenamefont {Zak}(1989)}]{zak_berrys_1989}%
  \BibitemOpen
\bibfield  {journal} {  }\bibfield  {author} {\bibinfo {author} {\bibfnamefont
  {J.}~\bibnamefont {Zak}},\ }\bibfield  {title} {\bibinfo {title}
  {Berry{\textquoteright}s phase for energy bands in solids},\ }\href
  {https://link.aps.org/doi/10.1103/PhysRevLett.62.2747} {\bibfield  {journal}
  {\bibinfo  {journal} {Phys. Rev. Lett.}\ }\textbf {\bibinfo {volume} {62}},\
  \bibinfo {pages} {2747} (\bibinfo {year} {1989})}\BibitemShut {NoStop}%
\bibitem [{\citenamefont {Halperin}(1987)}]{Halperin_1987}%
  \BibitemOpen
  \bibfield  {author} {\bibinfo {author} {\bibfnamefont {B.~I.}\ \bibnamefont
  {Halperin}},\ }\bibfield  {title} {\bibinfo {title} {Possible states for a
  three-dimensional electron gas in a strong magnetic field},\ }\href
  {https://doi.org/10.7567/jjaps.26s3.1913} {\bibfield  {journal} {\bibinfo
  {journal} {Jap. J Appl. Phys.}\ }\textbf {\bibinfo {volume} {26}},\ \bibinfo
  {pages} {1913} (\bibinfo {year} {1987})}\BibitemShut {NoStop}%
\bibitem [{\citenamefont {Tang}\ \emph {et~al.}(2019)\citenamefont {Tang},
  \citenamefont {Ren}, \citenamefont {Wang}, \citenamefont {Zhong},
  \citenamefont {Schneeloch}, \citenamefont {Yang}, \citenamefont {Yang},
  \citenamefont {Lee}, \citenamefont {Gu}, \citenamefont {Qiao},\ and\
  \citenamefont {Zhang}}]{TangNatureHall_2019}%
  \BibitemOpen
  \bibfield  {author} {\bibinfo {author} {\bibfnamefont {F.}~\bibnamefont
  {Tang}}, \bibinfo {author} {\bibfnamefont {Y.}~\bibnamefont {Ren}}, \bibinfo
  {author} {\bibfnamefont {P.}~\bibnamefont {Wang}}, \bibinfo {author}
  {\bibfnamefont {R.}~\bibnamefont {Zhong}}, \bibinfo {author} {\bibfnamefont
  {J.}~\bibnamefont {Schneeloch}}, \bibinfo {author} {\bibfnamefont {S.~A.}\
  \bibnamefont {Yang}}, \bibinfo {author} {\bibfnamefont {K.}~\bibnamefont
  {Yang}}, \bibinfo {author} {\bibfnamefont {P.~A.}\ \bibnamefont {Lee}},
  \bibinfo {author} {\bibfnamefont {G.}~\bibnamefont {Gu}}, \bibinfo {author}
  {\bibfnamefont {Z.}~\bibnamefont {Qiao}},\ and\ \bibinfo {author}
  {\bibfnamefont {L.}~\bibnamefont {Zhang}},\ }\bibfield  {title} {\bibinfo
  {title} {Three-dimensional quantum hall effect and metal–insulator
  transition in {ZrTe$_5$}},\ }\href
  {https://doi.org/10.1038/s41586-019-1180-9} {\bibfield  {journal} {\bibinfo
  {journal} {Nature}\ }\textbf {\bibinfo {volume} {569}},\ \bibinfo {pages}
  {537} (\bibinfo {year} {2019})}\BibitemShut {NoStop}%
\bibitem [{\citenamefont {Lv}\ \emph {et~al.}(2019)\citenamefont {Lv},
  \citenamefont {Qian},\ and\ \citenamefont {Ding}}]{HongNRP_2019}%
  \BibitemOpen
  \bibfield  {author} {\bibinfo {author} {\bibfnamefont {B.}~\bibnamefont
  {Lv}}, \bibinfo {author} {\bibfnamefont {T.}~\bibnamefont {Qian}},\ and\
  \bibinfo {author} {\bibfnamefont {H.}~\bibnamefont {Ding}},\ }\bibfield
  {title} {\bibinfo {title} {Angle-resolved photoemission spectroscopy and its
  application to topological materials},\ }\href
  {https://doi.org/10.1038/s42254-019-0088-5} {\bibfield  {journal} {\bibinfo
  {journal} {Nat. Rev. Phys.}\ }\textbf {\bibinfo {volume} {1}},\ \bibinfo
  {pages} {609} (\bibinfo {year} {2019})}\BibitemShut {NoStop}%
\bibitem [{\citenamefont {Zheng}\ \emph {et~al.}(2016)\citenamefont {Zheng},
  \citenamefont {Xu}, \citenamefont {Bian}, \citenamefont {Guo}, \citenamefont
  {Chang}, \citenamefont {Sanchez}, \citenamefont {Belopolski}, \citenamefont
  {Lee}, \citenamefont {Huang}, \citenamefont {Zhang}, \citenamefont {Sankar},
  \citenamefont {Alidoust}, \citenamefont {Chang}, \citenamefont {Wu},
  \citenamefont {Neupert}, \citenamefont {Chou}, \citenamefont {Jeng},
  \citenamefont {Yao}, \citenamefont {Bansil}, \citenamefont {Jia},
  \citenamefont {Lin},\ and\ \citenamefont {Hasan}}]{Zheng_ACS_2016}%
  \BibitemOpen
  \bibfield  {author} {\bibinfo {author} {\bibfnamefont {H.}~\bibnamefont
  {Zheng}}, \bibinfo {author} {\bibfnamefont {S.-Y.}\ \bibnamefont {Xu}},
  \bibinfo {author} {\bibfnamefont {G.}~\bibnamefont {Bian}}, \bibinfo {author}
  {\bibfnamefont {C.}~\bibnamefont {Guo}}, \bibinfo {author} {\bibfnamefont
  {G.}~\bibnamefont {Chang}}, \bibinfo {author} {\bibfnamefont {D.~S.}\
  \bibnamefont {Sanchez}}, \bibinfo {author} {\bibfnamefont {I.}~\bibnamefont
  {Belopolski}}, \bibinfo {author} {\bibfnamefont {C.-C.}\ \bibnamefont {Lee}},
  \bibinfo {author} {\bibfnamefont {S.-M.}\ \bibnamefont {Huang}}, \bibinfo
  {author} {\bibfnamefont {X.}~\bibnamefont {Zhang}}, \bibinfo {author}
  {\bibfnamefont {R.}~\bibnamefont {Sankar}}, \bibinfo {author} {\bibfnamefont
  {N.}~\bibnamefont {Alidoust}}, \bibinfo {author} {\bibfnamefont {T.-R.}\
  \bibnamefont {Chang}}, \bibinfo {author} {\bibfnamefont {F.}~\bibnamefont
  {Wu}}, \bibinfo {author} {\bibfnamefont {T.}~\bibnamefont {Neupert}},
  \bibinfo {author} {\bibfnamefont {F.}~\bibnamefont {Chou}}, \bibinfo {author}
  {\bibfnamefont {H.-T.}\ \bibnamefont {Jeng}}, \bibinfo {author}
  {\bibfnamefont {N.}~\bibnamefont {Yao}}, \bibinfo {author} {\bibfnamefont
  {A.}~\bibnamefont {Bansil}}, \bibinfo {author} {\bibfnamefont
  {S.}~\bibnamefont {Jia}}, \bibinfo {author} {\bibfnamefont {H.}~\bibnamefont
  {Lin}},\ and\ \bibinfo {author} {\bibfnamefont {M.~Z.}\ \bibnamefont
  {Hasan}},\ }\bibfield  {title} {\bibinfo {title} {Atomic-scale visualization
  of quantum interference on a {Weyl} semimetal surface by scanning tunneling
  microscopy},\ }\href {https://doi.org/10.1021/acsnano.5b06807} {\bibfield
  {journal} {\bibinfo  {journal} {ACS Nano}\ }\textbf {\bibinfo {volume}
  {10}},\ \bibinfo {pages} {1378} (\bibinfo {year} {2016})}\BibitemShut
  {NoStop}%
\end{thebibliography}%

\end{document}